\documentclass[journal]{IEEEtran}

\usepackage{lineno} 

\usepackage{amssymb} 
\usepackage{amsmath,amsfonts}
\usepackage{array}
\usepackage[caption=false,font=scriptsize,labelfont=sf,textfont=sf]{subfig}
\usepackage{textcomp}
\usepackage{stfloats}
\usepackage{url}
\usepackage{verbatim}
\usepackage{graphicx}
\usepackage{cite}

\usepackage[bookmarksnumbered=true]{hyperref}
\hypersetup{
hidelinks,
colorlinks=true,
linkcolor=blue,
citecolor=blue,
urlcolor=black
}
\usepackage[linesnumbered,ruled,vlined]{algorithm2e}
\usepackage{bbm} 

\usepackage{amsthm} 

\usepackage{hyperref} 
\usepackage{xcolor}

\usepackage{enumerate} 

\begin{document}

\newcounter{probcounter}
\newcommand{\refporbcounter}[1]{\refstepcounter{probcounter}\theprobcounter \label{#1}} 


\title{Graph Diffusion-Based AeBS Deployment and Resource Allocation in RSMA-Enabled URLLC Low-Altitude Wireless Networks}

\author{
Xudong Wang, 
Lei Feng,~\IEEEmembership{Member,~IEEE,}
Jiacheng Wang,
Hongyang Du,
Changyuan Zhao,\\
Wenjing Li,~\IEEEmembership{Member,~IEEE,}
and Ping Zhang,~\IEEEmembership{Fellow,~IEEE}

\thanks{\textit{Corresponding author: Lei Feng.}}
\thanks{Xudong Wang, Lei Feng, Wenjing Li and Ping Zhang are with the State Key Laboratory of Networking and Switching Technology, Beijing University of Posts and Telecommunications, Beijing 100876, China (e-mail: xdwang@bupt.edu.cn; fenglei@bupt.edu.cn; wjli@bupt.edu.cn; pzhang@bupt.edu.cn).}
\thanks{Jiacheng Wang and Changyuan Zhao are with the College of Computing and Data Science, Nanyang Technological University, Singapore (e-mail: jiacheng.wang@ntu.edu.sg; zhao0441@e.ntu.edu.sg).}
\thanks{Hongyang Du is with the Department of Electrical and Electronic Engineering, University of Hong Kong, Pok Fu Lam, Hong Kong SAR, China (e-mail: duhy@eee.hku.hk).}
}



\maketitle

\begin{abstract}
As a key component of low-altitude wireless networks, aerial base stations (AeBSs) provide flexible and reliable wireless coverage to support 6G ultra-reliable and low-latency communication (URLLC) services. However, limited spectrum resources and severe co-channel interference pose significant challenges to the deployment and resource allocation of AeBSs. To address these limitations, this paper proposes a novel rate-splitting multiple access (RSMA)-enabled transmission design to manage interference and enhance URLLC services in spectrum-constrained multi-AeBS networks. We formulate a joint optimization problem involving AeBS deployment, user association, and resource allocation to maximize the sum rate and coverage of system. Given the NP-hard nature of the problem, we propose a novel alternating optimization framework based on the generative graph diffusion models. Specifically, we model AeBSs and ground users as graph nodes, then we employ a discrete graph generation process solved via denoising diffusion to explore the combinatorial space of deployment and association strategies. Moreover, the successive convex approximation (SCA) is adopted to optimize AeBS beamforming and RSMA rate allocation under finite blocklength constraints. Extensive simulations demonstrate that the proposed algorithm outperforms existing methods in terms of convergence speed, sum rate, and coverage, while also exhibiting robust performance under varying network densities and interference levels.
\end{abstract}


\section{Introduction}
With the advent of sixth-generation (6G) wireless communication systems, ultra-reliable and low-latency communications (URLLC) have become a key service category to support mission-critical applications such as autonomous driving, industrial automation, and remote healthcare~\cite{10453965}. Although 6G URLLC aims to achieve sub-millisecond latency and ultra-high reliability, current terrestrial networks often struggle with coverage gaps, congestion, and limited flexibility, especially in dynamic or remote scenarios~\cite{10415249}. To overcome these limitations, low-altitude wireless networks (LAWNs) have emerged as a promising solution by leveraging aerial platforms to enhance connectivity~\cite{10906066}. LAWNs are typically composed of lightweight and rapidly deployable aerial infrastructures, enabling agile and on-demand connectivity services in scenarios where traditional terrestrial networks fall short~\cite{11045436}. A key component of LAWNs is the deployment of aerial base stations (AeBSs), which operate at low altitudes and serve as dynamic network nodes. AeBSs can be mounted on unmanned aerial vehicles (UAVs), balloons, or other aerial platforms. 

By establishing line-of-sight (LoS) links and rapidly reconfiguring the network in response to user mobility or network failures~\cite{9635653}, AeBSs are considered promising enablers for supporting 6G URLLC services in complex wireless environments. 
Some studies~\cite{8247211,10778659} have explored joint AeBS/UAV deployment and resource optimization for supporting URLLC services. 
Although AeBSs offer enhanced coverage and flexibility, the fact that orthogonal multiple access (OMA) schemes allocate distinct resources to avoid interference makes it challenging to further improve spectral efficiency and scalability for URLLC services, especially in interference-intensive LAWNs assisted by multiple AeBSs.
To overcome these challenges, non-orthogonal multiple access (NOMA) has been introduced as a crucial multiple access strategy in LAWNs. Unlike conventional orthogonal schemes, NOMA enables multiple users to simultaneously utilize the same time-frequency resources through power-domain multiplexing, thereby significantly improving spectral efficiency and supporting massive connectivity~\cite{7973036}. However, the performance of NOMA is often limited by strict user pairing requirements, sensitivity to channel state information, and the complexity of successive interference cancellation (SIC), particularly in dynamic LAWNs~\cite{9831440}.

Recently, rate-splitting multiple access (RSMA) has emerged as a more flexible and robust alternative. The core idea of RSMA is to split each user's message into a common part and a private part, enabling flexible interference management by partially decoding interference and partially treating it as noise~\cite{10813419}. This allows RSMA to effectively balance the trade-off between decoding complexity and system throughput, while being less sensitive to channel imperfections and user distributions. 
This approach enhances spectral efficiency, supports diverse quality-of-service requirements, and offers better resilience to channel variations. When applied to low-altitude wireless networks, RSMA provides significant benefits in managing interference and optimizing resource allocation, thereby further strengthening the communication capabilities of 6G URLLC systems~\cite{10411132}.
The authors in~\cite{9258414} first integrated RSMA into a UAV network, demonstrating the efficiency and application potential of RSMA in air-to-ground communications with infinite blocklength. The authors in~\cite{9953049} then extended it to finite blocklength, analyzing outage probability and error rate to prove the theoretical support for URLLC in multi-AeBS RSMA networks. Although showing potential, integrating RSMA into low-altitude wireless networks still faces several challenges.

On the one hand, most existing studies formulate the joint deployment and resource allocation problems in LAWNs as non-convex and NP-hard, making it challenging for traditional optimization methods to effectively solve them in dynamic environments~\cite{10529221}. Deep reinforcement learning (DRL) has been introduced as a data-driven alternative that can adapt to changing environments through trial-and-error interaction with the network. 
However, DRL easily suffers from slow convergence speed and poor generalization to unseen network states. To further improve solution efficiency and avoid local optima, generative diffusion models (GDMs) have recently been widely applied in network performance optimization~\cite{10852212}. Inspired by nonequilibrium thermodynamics, GDMs generate high-quality policies by iteratively denoising through a learned reverse diffusion process~\cite{10557650}. Compared to traditional optimization and DRL approaches, GDMs can provide improved generalization to diverse network conditions, greater robustness to model uncertainty, and better solution diversity, which is essential for escaping local optima in non-convex search spaces~\cite{10839238}. 
On the other hand, current research on RSMA mechanisms in low-altitude wireless networks remains limited, particularly in multi-AeBS networks where co-channel interference is considered. 
Given the topological characteristics of wireless networks, graph-based modeling and optimization can effectively capture the complex features of multi-AeBS networks. 
In~\cite{liu2024graph,wang2024generative}, the authors proposed the graph diffusion model-based approach to generate graph structures that meet specific optimization objectives. Benefiting from the generative nature, the graph-based diffusion models can globally model and efficiently explore the solution space. Therefore, incorporating graph diffusion models into RSMA-enabled multi-AeBS networks, which exhibit complex structures and high interference density, is a promising direction worth exploring.

Motivated by severe co-channel interference in multi-AeBS networks, this work adopts RSMA technology to mitigate interference and support URLLC connectivity through flexible resource allocation and UAV placement. To the best of our knowledge, this is the first study to investigate RSMA-enabled low-altitude wireless networks for URLLC services under co-channel interference. Additionally, a generative graph diffusion model is introduced to address the high complexity and suboptimal performance of traditional decision-based AI methods, effectively addressing dynamic deployment challenges in interference-prone low-altitude wireless networks. The main contributions of this study are summarized below: 
\begin{itemize}

    \item We propose a novel RSMA-enabled low-altitude wireless network framework, in which multiple AeBSs sharing the same frequency band are deployed to support ground users (GUs) with URLLC services. Then we formulate an optimization problem to jointly optimize AeBS deployment, user association, and resource allocation, with the objective of maximizing the achievable sum rate and coverage rate, while considering URLLC rate requirements and AeBS power constraints.
    \item We develop a novel \textbf{j}oint \textbf{s}uccessive convex approximation and \textbf{g}raph \textbf{d}iffusion (JSGD) optimization framework to solve this challenging mixed-integer nonlinear programming (MINLP) problem, which distinguishes our work from existing studies. Specifically, we transform the AeBS deployment and user association subproblem into a graph optimization model, where AeBSs and GUs are treated as nodes, and their associations are viewed as potential edges. Leveraging a graph diffusion architecture, JSGD effectively optimizes the joint deployment and association strategy using a denoising process. Furthermore, we adopt the successive convex approximation (SCA) algorithm for joint beamforming and rate allocation to solve the resource allocation subproblem.
    \item Simulation results validate the effectiveness of the proposed algorithm, demonstrating superior performance over existing decision-based AI baselines in terms of achievable sum rate, coverage rate, and convergence speed. Unlike traditional baselines that lack adequate exploration and global information utilization, JSGD exhibits excellent exploration capabilities through the denoising process, resulting in higher rewards. Furthermore, integrating JSGD with RSMA significantly outperforms conventional SDMA and NOMA schemes in terms of sum rate and coverage rate. 
\end{itemize}

The remainder of the paper is organized as follows. Section~\ref{sec_model} describes the system model and formulates the corresponding optimization problem. Section~\ref{algorithm} presents the proposed optimization algorithm. Section~\ref{sec_simulation} conducts the simulation experiments and analyzes the numerical results. Finally, Section~\ref{sec_conclusion} concludes the content of the paper.

\textit{Notations:}
Vectors and matrices are denoted using bold lowercase and bold uppercase letters, respectively. The operation $\mathbf{z}^{T}$, $\mathbf{z}^{H}$, and $\mathbf{z}^{-1}$ represent the transpose, conjugate transpose, and inversion of $\mathbf{z}$. The additive white Gaussian noise (AWGN) with mean $\mu$ and variance $\sigma^{2}$ is denoted by $\mathcal{C N}\left(\mu, \sigma^{2}\right)$. The function $Q\left ( z  \right )$ defines $Q(z) \triangleq \frac{1}{\sqrt{2 \pi}} \int_{z}^{\infty} \exp \left(-\frac{x^{2}}{2}\right) d x$. $\mathbb{C}$ denotes the set of complex numbers, and $\mathbb{E}\left\{\cdot\right\}$ stands for the expectation operator.

\section{System Model}\label{sec_model}

\begin{figure}[!t]
\centering
\includegraphics[width=0.41\paperwidth]{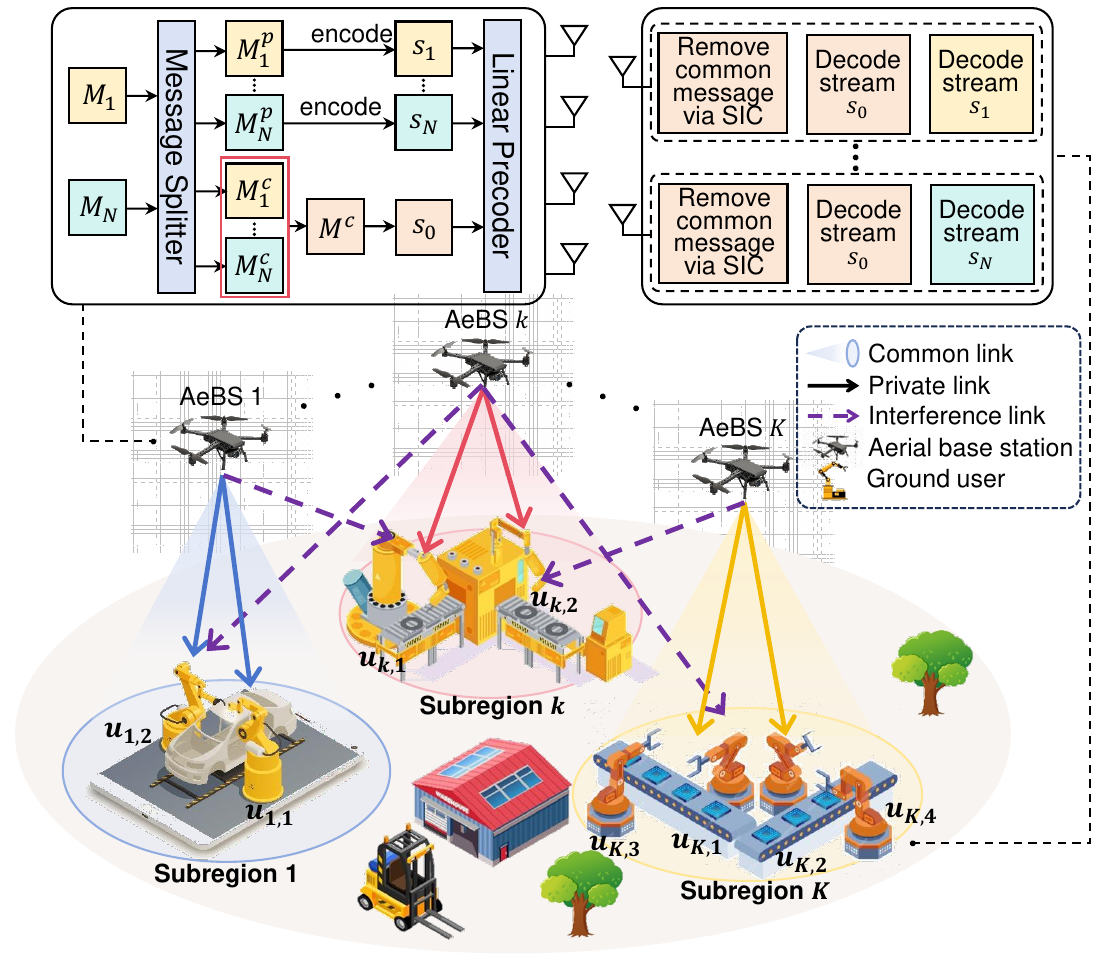}
\caption{RSMA-enabled URLLC low-altitude wireless networks.}
\label{fig:system_model}
\end{figure}

In this paper, we consider deploying multi-AeBS in RSMA-aided LAWNs to provide communication services to GUs that are randomly distributed. As illustrated in Fig.~\ref{fig:system_model}, the network model comprises a set of $N$ single-antenna GUs and $K$ UAVs equipped with $N_t$ antennas, where GUs are randomly distributed within a $L \times L \ \mathrm{m}^2$ task areas. The $K$ highly mobile and flexible UAVs are deployed as AeBSs to provide effective communication coverage and services to GUs in the dynamic scenario. Similar to~\cite{10411132}, we assume that the channel state information (CSI) is perfectly known at both the transmitters and receivers, and GUs are quasi-static within the considered task area. Let $\mathcal{K} = \left\{1, 2, \ldots, K\right\}$ represent the set of AeBSs indices, while $\mathcal{N} = \left\{1, 2, \ldots, N\right\}$ denotes the set of GUs indices. The notation $u_{k, n}$ represents the $n$-th GU served by $k$-th AeBS. We define the GU association variable $\alpha_{k,n} \in \{0,1\}$, where $\alpha_{k,n}=1$ indicates that GU $n \in \mathcal{N}$ is associated with the $k$-th AeBS, and $\alpha_{k,n}=0$ otherwise. Thus, the set of GUs served by $k$-th AeBS is defined as $\mathcal{N}_k\triangleq\{n\in\mathcal{N}\mid\alpha_{k,n}=1\}$, where $N_k=|\mathcal{N}_k|$ denotes the number of GUs served by $k$-th AeBS. We assume that each GU is exclusively linked to a single AeBS, i.e., $\mathcal{N}_{i} \cap \mathcal{N}_{j} = \emptyset, \forall i \ne j$. Consequently, it follows that the total number of GUs satisfies $N \ge {\textstyle \sum_{k \in \mathcal{K}}N_k}$.

\subsection{Network Model}

In the RSMA-enabled LAWNs, we aim to deploy multi-AeBS to provide communication services for GUs. Without compromising optimality, the position of the $k$-th AeBS is specified as $L_{ k}^\mathrm{b} = \left[x_{ k}^\mathrm{b}, y_{ k}^\mathrm{b}, H \right]$~\cite{9465671}. Besides, we consider that all AeBSs operate within a bounded area defined by $\left[X_{\mathrm{min}}, X_{\mathrm{max}}\right] \times \left[Y_{\mathrm{min}}, Y_{\mathrm{max}}\right]$. Thus, for each AeBS $k$ we have
\begin{align}
X_{\mathrm{min}} &\leq x_{k}^{b} \leq X_{\mathrm{max}}, \label{x_region} \\
Y_{\mathrm{min}} &\leq y_{k}^{b} \leq Y_{\mathrm{max}}. \label{y_region}
\end{align}
Moreover, to avoid potential collisions, a minimum separation distance denoted by $d_{\mathrm{min}}$ must be preserved between any two AeBSs. Accordingly, for $k$-th AeBS and $k^{'}$-th AeBS ($k \ne k^{'}, k, k^{'} \in \mathcal{K}$), the constraint is given by
\begin{equation}
\|L_k^b - L_{k^{'}}^b \|\geq d_{\mathrm{min}}.
\end{equation}

The position of GU $u_{k,n}$ is set as $L_{k, n}^\mathrm{u} = \left[x_{k, n}^\mathrm{u}, y_{k, n}^\mathrm{u}, 0\right]$. Thus, the distance from $k$-th AeBS to GU $u_{k,n}$ is expressed as
\begin{equation}
d_{k, n}=\sqrt{H^2+ l_{k,n}^2},
\end{equation}
where $l_{k,n} = \sqrt{\left(x_{k}^{\mathrm{b}}-x_{k, n}^{\mathrm{u}} \right)^{2}+\left(y_{k}^{\mathrm{b}}-y_{k, n}^{\mathrm{u}}\right)^{2}}$ represents the horizontal distance between $k$-th AeBS and $n$-th GU.

Due to the random distribution of GUs, they are susceptible to significant co-channel interference from intra-AeBS and inter-AeBS. We consider using one layer rate-splitting multiple access system where each GU is only served by one AeBS. In addition, each AeBS simultaneously supports at least two GUs. Thus, the constraints on $\alpha_{k,n}$ are formulated as
\begin{align}
&\sum_{k\in\mathcal{K}}\alpha_{k, n} =1,\quad\forall n \in \mathcal{N}, \\
&\sum_{n\in\mathcal{N}}\alpha_{k, n} \ge 2,\quad\forall k\in\mathcal{K}, \\
&\alpha_{k,n}\in \left\{0, 1\right\}, \quad \forall k\in\mathcal{K}, \forall n\in\mathcal{N}.
\end{align}
In addition, we consider that each AeBS possesses a communication radius $R$, within which it is capable of serving a cluster of GUs. Therefore, the association variables must comply with the constraints imposed by the communication range of the corresponding AeBS. In particular, the $n$-th GU is eligible to be served by the $j$-th AeBS only when it resides within the distance $R$, which is expressed as
\begin{equation}\label{communication_range}
d_{k,n}\leq R+S(1-\alpha_{k,n}),\quad \forall k\in\mathcal{K}, \forall n\in\mathcal{N},
\end{equation}
where $S$ is a very large positive number to ensure that constraint (\ref{communication_range}) is satisfied, indicating that if $\alpha_{k,n}=1$, the distance between the $k$-th AeBS and the $n$-th GU must not exceed the communication radius of the $k$-th AeBS.

To ensure the effectiveness of LAWNs, multi-AeBS networks must be meticulously planned to enhance coverage rate of system. The system-wide coverage rate is defined as
\begin{equation}
C = \frac{1}{N} \sum_{k \in \mathcal{K}}\sum_{n \in \mathcal{N}}\alpha_{k,n}\in \left[0,1\right].
\end{equation}

\subsection{Communication Model}

When AeBSs provide communication services to GUs, considering the air-to-ground (A2G) propagation channel, it is necessary to account for line-of-sight (LoS) channels as well as non-LoS (NLOS) channels, which introduce multiple signal reflection. Due to the additional losses caused by signal reflection from obstacles, the loss in the A2G channel model generally consists of free-space path loss and additional loss. When the signal undergoes LoS and NLoS propagation, the path losses are represented as $L_{k,n}^{\mathrm{LoS}}$ and $L_{k,n}^{\mathrm{NLoS}}$ respectively, which are given as follows~\cite{7510820}
\begin{align}
    L_{k,n}^{\mathrm{LoS}} &= 20\log\left(\frac{4\pi f_c d_{k,n}}{c}\right)+\zeta_{\mathrm{LoS}}, \\
    L_{k,n}^{\mathrm{NLoS}} &= 20\log\left(\frac{4\pi f_c d_{k,n}}{c}\right)+\zeta_{\mathrm{NLoS}},
\end{align}
where $f_c$ represents the carrier frequency of the signal, $c$ represents the speed of light, $\zeta_{\mathrm{NLoS}}$ and $\zeta_{\mathrm{LoS}}$ respectively represent the mean additional path losses of NLoS and LoS links, with their specific values determined by environmental conditions~\cite{7510820}. 
Therefore, the widely adopted A2G path loss model in~\cite{10778659} is considered, which is shown as
\begin{equation}
    PL_{k,n} =L_{k,n}^{\mathrm{LoS}} \times P_{k,n}^{\mathrm{LoS}} + L_{k,n}^{\mathrm{NLoS}} \times P_{k,n}^{\mathrm{NLoS}},
\end{equation}
where $P_{k,n}^{\mathrm{NLoS}}$ and $P_{k,n}^{\mathrm{LoS}}$ denote the propagation probabilities for NLoS and LoS propagation conditions, shown as~\cite{10778659}
\begin{align}
P_{k,n}^{\mathrm{LoS}} &= \frac{1}{1+\eta\exp\left(-\varsigma(\theta_{k,n} -\eta)\right)}, \\
P_{k,n}^{\mathrm{NLoS}} &= 1 - P_{k,n}^{\mathrm{LoS}},
\end{align}
where $\eta$ and $\varsigma$ denote the parameters determined by environmental characteristics, and $\theta_{k,n} = \mathrm{arctan} (H / l_{k,n})$ is the elevation angle between $k$-th AeBS and GU $u_{k,n}$. 

Considering the downlink transmission from AeBSs to GUs, the AeBSs use one layer RSMA transmission to provide services for GUs. At the $k$-th AeBS, the message $W_{k,n}$ of GU $u_{k,n}$ is partitioned into common and private data, i.e., $W_{k,n} = \left\{W_{k,n}^{c}, W_{k,n}^{p}\right\}$ for $\forall k \in \mathcal{K}, \forall n \in \mathcal{N}_k$. Next, all common data parts for the GUs in set $\mathcal{N}_k$ are jointly encoded into a single common stream $x_{k,0}$, which is decoded by each GU and subsequently removed via SIC to mitigate interference. This common stream $x_{k,0}$ is intended to be decoded by every GU in $\mathcal{N}_k$. And each GU's private message is encoded into its own private stream $x_{k,n}, \forall n \in \mathcal{N}_{k}$, which will be decoded only by its intended GU. Hence, the transmitted signal vector at $k$-th AeBS is $\mathbf{x}_{k} = \left[x_{k,0}, x_{k,1}, x_{k,2}, \ldots, x_{k,N_{k}}\right]^T$, and without loss of generality we assume $\mathbb{E}\left\{\mathbf{x}_{k} \mathbf{x}_{k}^{H}\right\} = \mathbf{I}$. To mitigate multi-user interference, we apply a precoding matrix $\mathbf{P}_{k} = \left[\mathbf{p}_{k,0}, \mathbf{p}_{k,1}, \mathbf{p}_{k,2}, \ldots, \mathbf{p}_{k,N_k}\right] \in \mathbb{C}^{N_{t}\times \left(N_k +1\right)}$, which assigns appropriate weights to each stream across the $N_t$ antennas. Finally, the precoded streams are superimposed and transmitted as $\mathbf{s}_{k} = \left[s_{k,1}, \ldots, s_{k,N_t}\right]^{T} = \mathbf{P}_{k} \mathbf{x}_{k}$. The transmit signal $\mathbf{s}_{k}$ is subject to a power budget $P_{\mathrm{max}}$, i.e., $\mathbb{E}\left\{\mathbf{s}_{k}^{H} \mathbf{s}_{k}\right\} \le P_{\mathrm{max}}, \forall k \in \mathcal{K}$, which is shown as
\begin{equation}
    \sum_{n = 0}^{N_k} \left\|\mathbf{p}_{k,n} \right\|^2\leq P_{\mathrm{max}}.
\end{equation}
The received signal at GU $u_{k,n}$ is then obtained as
\begin{equation}
    y_{k,n} = \mathbf{h}_{k,n}^{H} \mathbf{P}_{k} \mathbf{x}_{k} + \sum_{i \in \mathcal{K} \setminus \left\{k\right\} }^{}\mathbf{h}_{i,n}^{H} \mathbf{P}_{i} \mathbf{x}_{i} + n_{k,n},
\end{equation}
where $n_{k,n} \sim \mathcal{C N}\left(0, \sigma^{2}\right)$, and $\mathbf{h}_{k,n} = \left[h_{k,n}^{1}, \ldots, h_{k,n}^{N_t}\right]^{T} \in \mathbb{C}^{N_t \times 1}$ is the A2G channel from the $k$-th AeBS to GU $u_{k,n}$, and each channel coefficient $h_{k, n}^{j}$ can be expressed as
\begin{equation}
    h_{k,n}^{j} = 10^{-\frac{PL_{k,n}}{20}}, \quad \forall j \in \left\{1, \dots, N_t\right\}.
\end{equation}

Based on \cite{8846706}, we assume the channels are perfectly known at both transmitters and receivers. The decoding process at GU $u_{k,n}$ is outlined as follows. First, the common stream $x_{k, 0}$ is decoded by GU $u_{k,n}$, while treating all other signal components as noise. GU $u_{k,n}$ then extracts the information intended for it from this decoded common stream. Next, each GU adopts SIC to eliminate the influence of the decoded common stream, which improves the decoding of the private stream. Thereafter, GU $u_{k,n}$ decodes its own private stream $x_{k, n}$, treating the remaining private streams from other GUs as background noise. Consequently, we derive the received signal-to-interference-plus-noise ratios (SINRs) of the common stream $x_{k, 0}$ as
\begin{equation}
    \gamma_{k,n}^c =\frac{\alpha_{k,n} |\mathbf{h}_{k,n}^H \mathbf{p}_{k,0}|^2}{I_{\mathrm{in}}^{c} + I_{\mathrm{out}} +\sigma^2},
\end{equation}
where $I_{\mathrm{in}}^{c}$ represents the intra-AeBS interference, expressed as $I_{\mathrm{in}}^{c} = \sum_{j \in \mathcal{N}_k} \alpha_{k,j} |\mathbf{h}_{k,n}^H \mathbf{p}_{k,j}|^2$, while $I_{\mathrm{out}}$ is the co-channel interference power from other AeBSs, and it is given as
\begin{equation}
I_{\mathrm{out}} = \sum_{i \in \mathcal{K}\setminus \left\{k\right\}^{}} \left(|\mathbf{h}_{i,0}^{H} \mathbf{p}_{i,0}|^2 + \sum_{n \in \mathcal{N}} \alpha_{i,n} |\mathbf{h}_{i,n}^{H} \mathbf{p}_{i,n}|^2\right).
\end{equation}
To guarantee that the common stream is decodable by all GUs served by AeBS $k$, the achievable common rate $R_{k}^{c}$ must satisfy the following constraint~\cite{mao2018rate}:
\begin{equation}
R_k^c\leq\min_{n\in\mathcal{N}_k}c_{k,n},
\end{equation}
where
\begin{equation}
c_{k,n} = \log_{2}{\left(1 + \gamma_{k,n}^{c}\right)} -\sqrt{\frac{V(\gamma_{k,n}^{c})}{D^{c}}}\frac{Q^{-1}(\varepsilon)}{\ln2},
\end{equation}
represents the common rate that GU $u_{k,n}$ can reliably decode. Here, $V(x)=1-(1+x)^{-2}$ is the channel dispersion, $D^{c}$ is the transmission channel blocklength of the common message, and $\varepsilon$ represents the decoding error probability (DEP).
According to the principles of rate-splitting decoding, the common rate $R_k^c$ is a shared rate available to all GUs served by $k$-th AeBS. By decoding $r_{k,n}^{c}$ as the general common rate of GU $u_{k,n}$, we have
\begin{equation}
R_k^c=\sum_{n\in\mathcal{N}_k}r_{k,n}^c.
\end{equation}
Once the common stream has been removed, GU $u_{k,n}$ proceeds to decode its private stream utilizing the following SINR
\begin{equation}
    \gamma_{k,n}^p =\frac{\alpha_{k,n} |\mathbf{h}_{k,n}^H \mathbf{p}_{k,n}|^2}{I_{\mathrm{in}}^{p} + I_{\mathrm{out}} +\sigma^2},
\end{equation}
where $I_{\mathrm{in}}^{p} = \sum_{i \in \mathcal{N} \setminus \{ n \}} \alpha_{k,i} |\mathbf{h}_{k,n}^H \mathbf{p}_{k,i}|^2$ represents the residual intra-AeBS interference that remains after the common stream has been removed. Hence, the achievable private rate for GU $u_{k,n}$ can be obtained as
\begin{equation}
r_{k,n}^{p} = \log_{2}{\left(1 + \gamma_{k,n}^{p}\right)} -\sqrt{\frac{V(\gamma_{k,n}^{p})}{D^{p}}}\frac{Q^{-1}(\varepsilon)}{\ln2},
\end{equation}
where $D^{p}$ is the transmission channel blocklength of the private message.
Then, the total achievable rate for GU $u_{k,n}$ is obtained as
\begin{equation}
    R_{k,n} = r_{k,n}^{c} + r_{k,n}^{p}, \quad \forall k \in \mathcal{K}, n \in \mathcal{N}_k.
\end{equation}
Finally, the system's achievable sum rate is calculated as
\begin{equation}
    R_{\mathrm{sum}}=\sum_{k\in \mathcal{K}}\sum_{n\in \mathcal{N}}\alpha_{k,n} R_{k,n}.
\end{equation}

In URLLC systems, such as industrial cyber-physical systems, GUs have strict latency requirements. 
We can reformulate the latency requirement as a rate constraint. Specifically, if a data packet of $D$ bits must be delivered within a given latency bound $T_{th}$ seconds, the transmission rate must satisfy $r\geq\frac{D}{B T_{th}}$ (bit/s/Hz), where $B$ represents the avaliable bandwidth. Therefore, the rate of GU $n$ served by AeBS $k$ should be $R_{k,n} \ge R_{\mathrm{min}} = \frac{D^{c}+D^{p}}{B T_{th}}$ (bit/s/Hz), where $R_{\mathrm{min}}$ is the GU's minimum rate requirement.

\subsection{Problem Formulation}

Our goal is to jointly design AeBS deployment, user association and resource allocation scheme, with the objective of maximizing the overall system utility, which incorporates both the sum rate and the coverage rate. Enhancing the coverage enables the network to connect a greater number of GUs, while increasing sum rate accelerates information transfer in LAWNs, both of which enhance decision support and collaborative communications. We define the system utility as
\begin{equation} \label{utility_function}
    \mathcal{U}= \lambda_1 C+\lambda_2 \frac{R_{\mathrm{sum}}}{R_N},
\end{equation}
where $\lambda_{1}$ and $\lambda_{2}$ are weighting factors that balance the trade-off between maximizing coverage and sum rate, respectively, and $R_N$ denotes the normalization factor used to scale the sum-rate term so that coverage $C$ and sum rate $R_{\mathrm{sum}}$ are of comparable magnitude for optimization. This optimization problem is expressed as
\begin{subequations}
	\begin{align}
  \mathcal{P}0: 
  &\mathop{\mathrm{max}}\limits_{\boldsymbol{\alpha}, \mathbf{L}, \mathbf{P}, \mathbf{C}} \ \mathcal{U}
		\\
		\mathrm{s.t.} \ 
		\mathrm{C1:} &\sum_{k\in\mathcal{K}}\alpha_{k, n} =1, \forall n \in \mathcal{N}, \\
        \mathrm{C2:} &2 \leq \sum_{n\in\mathcal{N}_k}\alpha_{k, n} \leq N_{a}, \forall k\in\mathcal{K}, \\
        \mathrm{C3:} &\alpha_{k,n} \in \left\{0, 1\right\}, \forall n\in\mathcal{N}, \forall k\in\mathcal{K}, \\
        \mathrm{C4:} &(\ref{x_region}) - (\ref{y_region}), \forall k \in \mathcal{K}, \\
        \mathrm{C5:} & \sum_{n\in\mathcal{N}_k} r_{k,n}^{c} \leq \min_{n\in\mathcal{N}_k}c_{k,n}, \forall k \in \mathcal{K}, \label{generative_data_stream_can_be_decoded} \\
        \mathrm{C6:} & R_{k,n} \ge R_{\mathrm{min}}, \forall k \in \mathcal{K}, \forall n \in \mathcal{N}_k \label{minimum_achievable_sum_rate} \\
        \mathrm{C7:} & r_{k,n}^{c} \ge 0, \forall k \in \mathcal{K}, \forall n \in \mathcal{N} \label{nonnegativity_of_common_rate} \\
        \mathrm{C8:} &\sum_{n = 0}^{N_k} \left\|\mathbf{p}_{k,n} \right\|^2\leq P_{\mathrm{max}}, \forall k \in \mathcal{K}, \label{constraint_sum_of_transmit_power} \\
        \mathrm{C9:} &\|L_k^b - L_{k^{'}}^b \|\geq d_{\mathrm{min}}, k, k^{'} \in \mathcal{K}, k \ne k^{'}, \\
        \mathrm{C10:} &d_{k,n}\leq R+S(1-\alpha_{k,n}), \forall k\in\mathcal{K}, \forall n\in\mathcal{N}, \label{communication_range_constraint}
	\end{align}
\end{subequations}
where $\boldsymbol{\alpha} = \left\{\alpha_{k,n}| k \in \mathcal{K}, n \in \mathcal{N}\right\}$ represents user association policy, $\mathbf{L} = \left\{L_{k}^{b}| k \in \mathcal{K}\right\}$ denotes the coordinate of AeBSs, $\mathbf{P}= \left\{\mathbf{P}_k| k \in \mathcal{K}\right\}$ denotes the beamforming matrix, $\mathbf{C} = \left\{r_{k,n}^{c}| k \in \mathcal{K}, n \in \mathcal{N}_k\right\}$ denotes the common rate vector, $N_a$ denotes the upper bound to the amount of GUs one AeBS can serve. $R_{\mathrm{min}}$ denotes the GU's demand on the achievable rate. Constraint C1 ensures that each GU can be only served by one AeBS. C2 ensures each AeBS can serve at least two and at most $N_a$ GUs using RSMA principle. C3 imposes a Boolean constraint on the user association strategy. C4 ensures that the locations of AeBSs are confined within a fixed task area. C5 enforces the common stream must be decodable by all associated GUs. C6 is the transmission delay constraint. C7 guarantees the nonnegativity of common rate. C8 indicates the transmit power budget. C9 guarantees collision avoidance between any two AeBSs. C10 is the communication range constraint.

Considering the integer variable $\alpha_{j,n}$ and nonconvex functions of coverage rate and achievable sum rate, $\mathcal{P}0$ constitutes a mixed-integer nonlinear programming (MINLP) problem that is challenging to solve directly. Conventional optimization methods, such as reinforcement learning (RL), often suffer from slow convergence owing to their high computational complexity, thus a more efficient solution strategy is needed.

\section{Proposed JSGD Optimization Algorithm}\label{algorithm}

In the section, we propose a novel \textbf{j}oint \textbf{S}CA and \textbf{g}raph \textbf{d}iffusion (JSGD) optimization framework to solve problem $\mathcal{P}0$. We first decompose problem $\mathcal{P}0$ into two sub-problems, including sub-problem $\mathbf{\mathcal{P}1}$: beamforming and common rate control; and sub-problem $\mathbf{\mathcal{P}2}$: joint AeBS deployment and user association. We then employ an iterative algorithm to find the optimal solution of $\mathcal{P}0$. Specifically, we address AeBS deployment and user association using the proposed generative graph diffusion model (GGDM), and we tackle beamforming and common rate control using SCA method.

\subsection{SCA-based Resource Allocation for Sub-problem $\mathbf{\mathcal{P}1}$}

We first focus on dealing with the resource allocation sub-problem. With the given AeBS deployment $\mathbf{L}$ and user association $\boldsymbol{\alpha}$ between AeBSs and GUs, $\mathcal{P}0$ is rewritten as
\begin{subequations}\label{subproblem_1}
	\begin{align}
  \mathcal{P}1: 
  \mathop{\mathrm{max}}\limits_{\mathbf{P}, \mathbf{C}} \ 
		 & \kappa R_{\mathrm{sum}} \qquad \qquad \label{P1_objective}
		\\
		\mathrm{s.t.} \ 
		&\mathrm{C5}\sim \mathrm{C8},
	\end{align}
\end{subequations}
where $\kappa = \lambda_2 / R_N$ is a constant.
Owing to the non-convex rate expressions, the objective function (\ref{P1_objective}) and constraints (\ref{generative_data_stream_can_be_decoded}), (\ref{minimum_achievable_sum_rate}) and (\ref{nonnegativity_of_common_rate}) in $\mathcal{P}1$ are non-convex. To address this resource allocation sub-problem, we adopt the SCA algorithm to jointly optimize beamforming matrix $\mathbf{P}$ and common rate $\mathbf{C}$.

We begin by introducing the slack variable sets $\boldsymbol{\varphi} = \left\{\varphi_{k,n}^{p} | k \in \mathcal{K}, n \in \mathcal{N}_k \right\}$, $\boldsymbol{\nu}_c = \left\{\nu_{k,n}^{c} | k \in \mathcal{K}, n \in \mathcal{N}_k\right\}$, $\boldsymbol{\nu}_p = \left\{\nu_{k,n}^{p} | k \in \mathcal{K}, n \in \mathcal{N}_k\right\}$, $\boldsymbol{\chi}_c = \left\{\chi_{k,n}^{c} | k \in \mathcal{K}, n \in \mathcal{N}_k\right\}$, $\boldsymbol{\chi}_p = \left\{\chi_{k,n}^{p} | k \in \mathcal{K}, n \in \mathcal{N}_k\right\}$, 
where $\varphi_{k,n}^{p}$ denotes the minimum threshold for the private rate $r_{k,n}^{c}$. $\nu_{k,n}^{c}$ and $\nu_{k,n}^{p}$ represent the minimum required SINRs for the common and private streams, respectively. The parameters $\chi_{w,k}^{c}$ and $\chi_{w,k}^{p}$ correspond to the upper bounds on the interference-plus-noise terms associated with $\nu_{k,n}^{c}$ and $\nu_{k,n}^{p}$, respectively. Similarly, $\pi_{k,n}^{c}$ and $\pi_{k,n}^{p}$ specify the lower limits for the power of useful signals corresponding to $\nu_{k,n}^{c}$ and $\nu_{k,n}^{p}$, respectively. For a fixed DEP and blocklength, $Y =  \frac{Q^{-1}\left(\epsilon \right)} {\sqrt{l}} \log_{2}{e}$ is a constant. Moreover, we define $y_{k,n}^{i}=1-\left(1+\nu_{k,n}^{i}\right)^{-2}, \forall i \in \left \{ c,p \right \} $. Based on these introduced variables, sub-problem (\ref{subproblem_1}) is rewritten as
\begin{subequations} \label{subproblem_1-2}
\begin{align}
 \mathcal{P}1\textit{-}1: &\ \max _{\mathbf{P}, \mathbf{C}, \boldsymbol{\varphi}, \boldsymbol{\nu}_{c}, \atop \boldsymbol{\nu}_p, \boldsymbol{\chi_{c}}, \boldsymbol{\chi}_p} { \sum_{k\in\mathcal{K}} \sum_{n\in \mathcal{N}_k} \left(r_{k,n}^{c}+\varphi_{k,n}^{p}\right)} \\
    \text { s.t.} \ & \log_{2}\left(1+\nu_{k,n}^{c}\right) - Y\sqrt{y_{k,n}^{c}} \ge \sum_{n\in \mathcal{N}_k} r_{k,n}^{c}, \forall k, n \label{SCA_constraint_common_rate} \\
    &\log_{2}\left(1+\nu_{k,n}^{p}\right) - Y\sqrt{y_{k,n}^{p}} \ge  \varphi_{k,n}^{p}, \forall k, n, \label{SCA_constraint_private_rate} \\
    &\frac{ |\mathbf{h}_{k,n}^H \mathbf{p}_{k,0} |^2}{\chi_{k,n}^{c}} \ge \nu_{k,n}^{c}, \quad \forall k,n, \label{common_SINR_constraint} \\    
    &\frac{ |\mathbf{h}_{k,n}^H \mathbf{p}_{k,n} |^2}{\chi_{k,n}^{p}} \ge \nu_{k,n}^{p}, \quad \forall k,n, \label{private_SINR_constraint} \\ 
    & \chi_{k,n}^{c} \ge \sum_{j \in \mathcal{N}_k} |\mathbf{h}_{k,n}^H \mathbf{p}_{k,j} |^2+ \Psi_{k,n} + \sigma^{2}, \forall k,n, \label{SINR_approximate_common_rate} \\
    & \chi_{w,k}^{p} \ge \sum_{j \in \mathcal{N}_k \setminus n } |\mathbf{h}_{k,n}^H \mathbf{p}_{k,j} |^2+ \Psi_{k,n} + \sigma^{2}, \forall k,n, \label{SINR_approximate_private_rate} \\
    & r_{k,n}^{c} + \varphi_{k,n}^{p} \ge R_{\mathrm{min}}, \quad \forall k,n, \label{minimum_achievable_sum_rate_approximate} \\
    & (\textup{\ref{nonnegativity_of_common_rate}}), (\textup{\ref{constraint_sum_of_transmit_power}}), \notag
\end{align}
\end{subequations}
where $\Psi_{k,n} = \sum_{i \in \mathcal{K}\setminus \left\{k\right\}^{}} \left(|\mathbf{h}_{i,0}^{H} \mathbf{p}_{i,0}|^2 + \sum_{n \in \mathcal{N}_{i}^{}} |\mathbf{h}_{i,n}^{H} \mathbf{p}_{i,n}|^2\right)$. Problem (\ref{subproblem_1-2}) remains non-convex owing to the non-convexity of constraints (\ref{SCA_constraint_common_rate})-(\ref{private_SINR_constraint}). Applying the Taylor series expansion, constraints (\ref{SCA_constraint_common_rate}) and (\ref{SCA_constraint_private_rate}) at $\nu_{k,n}^{c\left[\ell\right]}$ and $\nu_{k,n}^{p\left[\ell\right]}$ at iteration $\ell$ are approximated as 
\begin{subequations} \label{SCA_Taylor_rate_1}
\begin{align}
& \log _{2}\left(1+\nu_{k,n}^{c}\right) - 
Y\Bigg\{\Bigg[ 1 - \left( 1 + \nu _ {k,c} ^ {c [ \ell ] } \right) ^ { - 2 } \Bigg] ^ { - \frac { 1 } { 2 } } 
\Bigg[\left(1+\nu_{k,n}^{c [\ell]}\right)^{-3} \notag \\
& \ \times \left(\nu_{k,n}^{c}-\nu_{k,n}^{c [\ell]}\right) - \left(1+\nu_{k,n}^{c [\ell]}\right)^{-2} + 1\Bigg]\Bigg\} \geq \sum_{j \in \mathcal{N}_k} r_{k,n}^{c}, \label{SCA_constraint_common_rate_1} \\
& \log _{2}\left(1+\nu_{k,n}^{p}\right) - 
Y\Bigg\{\Bigg[ 1 - \left( 1 + \nu _ {k,n} ^ {p [ \ell ] } \right) ^ { - 2 } \Bigg] ^ { - \frac { 1 } { 2 } } 
\Bigg[\left(1+\nu_{k,n}^{p [\ell]}\right)^{-3} \notag \\
& \ \times \left(\nu_{k,n}^{p}-\nu_{k,n}^{p [\ell]}\right) - \left(1+\nu_{k,n}^{p [\ell]}\right)^{-2} + 1\Bigg]\Bigg\} \geq \varphi_{k,n}^{p}. \label{SCA_constraint_private_rate_1}
\end{align}
\end{subequations}
Also, we separately approximate constraints (\ref{common_SINR_constraint}) and (\ref{private_SINR_constraint}) at the respective points $\left(\mathbf{p}_{k,0}^{\left[\ell\right]}, \chi_{k,n}^{c \left[\ell\right]}\right)$ and $\left(\mathbf{p}_{k,n}^{\left[\ell\right]}, \chi_{k,n}^{p \left[\ell\right]}\right)$, as detailed follows:
\begin{subequations} \label{SCA_Taylor_rate_2}
\begin{align}
& \ \frac{2\Re\left\{\left(\mathbf{p}_{k,0}^{[\ell]}\right)^H\mathbf{h}_{k,n}\mathbf{h}_{k,n}^H\mathbf{p}_{k,0}\right\}}{\chi_{k,n}^{c[\ell]}}-\frac{\left|\mathbf{h}_{k,n}^H\mathbf{p}_{k,0}^{\left[\ell \right]}\right|^2\chi_{k,n}^{c}}{\left(\chi_{k,n}^{c[\ell]}\right)^2}\geq\nu_{k,n}^{c}, \\
& \ \frac{2\Re\left\{\left(\mathbf{p}_{k,n}^{[\ell]}\right)^H\mathbf{h}_{k,n}\mathbf{h}_{k,n}^H\mathbf{p}_{k,n}\right\}}{\chi_{k,n}^{p[\ell]}}-\frac{\left|\mathbf{h}_{k,n}^H\mathbf{p}_{k,n}^{\left[\ell \right]}\right|^2\chi_{k,n}^{p}}{\left(\chi_{k,n}^{p[\ell]}\right)^2}\geq\nu_{k,n}^{p}.
\end{align}
\end{subequations}

By following the aforementioned processing steps, the original non-convex formulation is converted into a convex optimization problem, allowing it to be efficiently solved using the SCA algorithm. 
The core idea of the SCA method is to iteratively approximate the initial non-convex problem by a sequence of convex subproblems, each of which can be successively tackled using standard optimization techniques.
Based on the optimal result~$(\mathbf{P}^{\left[\ell\right]}, \boldsymbol{\nu}_{c}^{\left[\ell\right]}, \boldsymbol{\nu}_p^{\left[\ell\right]}, \boldsymbol{\chi}_c^{\left[\ell\right]}, \boldsymbol{\chi}_p^{\left[\ell\right]})$ obtained in the previous iteration $\ell-1$, the sub-problem at iteration $\ell$ can be solved as
\begin{subequations} \label{subproblem_1-3}
\begin{align}
    \mathcal{P}1\textit{-}2: \ \max _{\mathbf{P}, \mathbf{C}, \boldsymbol{\varphi}, \boldsymbol{\nu}_{c}, \atop \boldsymbol{\nu}_p, \boldsymbol{\chi_{c}}, \boldsymbol{\chi}_p} &{ \sum_{k\in\mathcal{K}} \sum_{n\in \mathcal{N}_k} \left(r_{k,n}^{c}+\varphi_{k,n}^{p}\right)} \qquad \quad \\
    \text { s.t.} \quad & (\textup{\ref{nonnegativity_of_common_rate}}), (\textup{\ref{constraint_sum_of_transmit_power}}), (\textup{\ref{SINR_approximate_common_rate}}), \notag \\ 
    & (\textup{\ref{SINR_approximate_private_rate}}),(\textup{\ref{minimum_achievable_sum_rate_approximate}}), (\textup{\ref{SCA_Taylor_rate_1}}), (\textup{\ref{SCA_Taylor_rate_2}}). \notag
\end{align}
\end{subequations}
For ease of notation, let $\rho = \sum_{k\in\mathcal{K}} \sum_{n\in \mathcal{N}_k} \left(r_{k,n}^{c}+\varphi_{k,n}^{p}\right)$, and $\rho^{\left[\ell\right]} = \sum_{k\in\mathcal{K}} \sum_{n\in \mathcal{N}_k} \left(r_{k,n}^{c \left[\ell\right]}+\varphi_{k,n}^{p \left[\ell\right]}\right)$. The detailed procedure for beamforming and common rate design is summarized in Algorithm \ref{SCA_algorithm}, where $\tau$ denotes the convergence tolerance. Since the solution in iteration $\ell-1$ remains feasible for iteration $\ell$, Algorithm \ref{SCA_algorithm} is guaranteed to converge. The convergence of the SCA method has been rigorously established in prior studies, such as \cite{palacios1982nonlinear}, and thus the proof in our paper is omitted here for brevity.

\begin{algorithm}[!t]
    \caption{Beamforming and Common Rate Control for RSMA-enabled LAWNs}
    \label{SCA_algorithm}
    \SetKwData{Left}{left}\SetKwData{This}{this}\SetKwData{Up}{up}
 \SetKwFunction{FindCompress}{FindCompress}
	\SetKwInOut{Input}{Input}\SetKwInOut{Output}{Output}
	
	\Input{the convergence tolerance $\tau$, the iteration index $\ell\gets 0$, $\rho^{\left[\ell\right]} \gets 0$, $\mathbf{P}^{\left[\ell\right]}$, $\boldsymbol{\nu}_{c}^{\left[\ell\right]}$, $\boldsymbol{\nu}_p^{\left[\ell\right]}$,$\boldsymbol{\chi}_c^{\left[\ell\right]}$, $\boldsymbol{\chi}_p^{\left[\ell\right]}$.}
        
    \Repeat{$\left | \rho^{\left[\ell\right]}-\rho^{\left[\ell-1\right]} \right | < \tau$}{
      $\ell\gets \ell+1$; \\      
      Solve problem (\ref{subproblem_1-3}) using $\mathbf{P}^{\left[\ell-1\right]}$, $\boldsymbol{\nu}_{c}^{\left[\ell-1\right]}$, $\boldsymbol{\nu}_p^{\left[\ell-1\right]}$, $\boldsymbol{\chi}_c^{\left[\ell-1\right]}$, $\boldsymbol{\chi}_p^{\left[\ell-1\right]}$; \\ 
      Record the optimal sum rate $\rho^{*}$ and the optimal solutions $\mathbf{P}^{*}$, $\boldsymbol{\nu}_{c}^{*}, \boldsymbol{\nu}_p^{*}$, $\boldsymbol{\chi}_c^{*}, \boldsymbol{\chi}_p^{*}$; \\
      Update $\rho^{\left[\ell\right]} \gets~\rho^{*}$, $\mathbf{P}^{\left[\ell\right]} \gets \mathbf{P}^{*}$, $\boldsymbol{\nu}_{c}^{\left[\ell\right]} \gets \boldsymbol{\nu}_{c}^{*}$, $\boldsymbol{\nu}_{p}^{\left[\ell\right]} \gets \boldsymbol{\nu}_p^{*}$, $\boldsymbol{\chi}_c^{\left[\ell\right]} \gets \boldsymbol{\chi}_c^{*}$, $\boldsymbol{\chi}_p^{\left[\ell\right]} \gets \boldsymbol{\chi}_p^{*}$;
    }

\Output{$\rho^{*}, \mathbf{C}^{*}, \mathbf{P}^{*}, \boldsymbol{\nu}_{c}^{*}, \boldsymbol{\nu}_p^{*}, \boldsymbol{\chi}_c^{*}$.}

Recovery~the~beamforming~strategy~from~$\mathbf{P}^{*}$~by~eigendecomposition~technique~or~Gaussian~randomization.
\end{algorithm}

\subsection{GGDM-based AeBS Deployment and User Association for Sub-problem $\mathbf{\mathcal{P}2}$}

For any given beamforming matrix $\mathbf{P}$ and common rate vector $\mathbf{C}$, the joint AeBS deployment and user association sub-problem $\mathcal{P}2$ is simplified as

\begin{subequations}\label{subproblem_2}
	\begin{align}
  \mathcal{P}2: 
  \mathop{\mathrm{max}}\limits_{\boldsymbol{\alpha}, \mathbf{L}} \ 
		 & \mathcal{U} 
		\\
		\mathrm{s.t.} \ 
		&\mathrm{C1}\sim \mathrm{C4}, \mathrm{C9}\sim \mathrm{C10}.
	\end{align}
\end{subequations}
Since AeBSs and GUs can be treated as graph nodes and their potential associations as edges, naturally recasting the joint deployment-association problem into a graph optimization task. Given the dynamic topology and high-dimensional resource constraints of RSMA-assisted LAWNs, conventional heuristics often get trapped in suboptimal solutions, while DRL approaches struggle to capture the complex high-dimensional features. In contrast, generative graph diffusion models can explore the solution space globally via iterative denoising, thereby avoiding local optima and mitigating training instability issues~\cite{liu2024graph}. Motivated by these advantages, we propose a graph diffusion model-based approach to solve sub-problem $\mathcal{P}2$.

\subsubsection{Graph Diffusion Stage}Drawing inspiration from non-equilibrium thermodynamics, graph diffusion models employ two sequential Markov processes, i.e., a forward diffusion process and a reverse denoising process, to generate new graph data~\cite{10419041,wang2024generative}. We represent the network state as a graph $\mathcal{G}=\left(\mathcal{V}, \mathcal{E}\right)$, where $\mathcal{V}$ denotes vertices set which represents AeBS deployment, and $\mathcal{E}$ denotes the corresponding adjacency matrix indicating user associations between AeBSs and GUs. Therefore, the sub-problem $\mathcal{P}2$ is equivalently transformed into finding an optimal initial $\mathcal{G}^{0}$ that satisfies all given conditions. Based on the discrete generative nature, we assume the nodes and edges can each take on $\left | \mathcal{V}  \right |$ and $\left | \mathcal{E}  \right |$ potential states, respectively, and we adopt one-hot encoding to embed these categorical features.

In the forward diffusion phase, the graph $\mathcal{G}^{0}$ is gradually perturbed into a random graph $\mathcal{G}^{T}$ over $T$ steps by sequentially adding noise. According to~\cite{vignac2022digress}, the diffusion is applied independently to each node and each edge. Let $Q_{\mathcal{V}}^{t}$ and $Q_{\mathcal{E}}^{t}$ be the transition matrices at the $t$-th diffusion step for vertices and edges, respectively. The transition probabilities for any vertex or edge follows a uniform or marginal distribution, which are defined as
\begin{subequations}
\begin{align}
[\boldsymbol{Q}_{\mathcal{V}}^{t}]_{ij}=p(\boldsymbol{v}^{t}=j|\boldsymbol{v}^{t-1}=i), \\
[\boldsymbol{Q}_{\mathcal{E}}^{t}]_{ij}=p(\boldsymbol{e}^{t}=j|\boldsymbol{e}^{t-1}=i).
\end{align}
\end{subequations}
Starting from the initial graph $\mathcal{G}^{0}=\left(\mathcal{V}^{0}, \mathcal{E}^{0}\right)$, the overall diffusion process is formulated as
\begin{equation}
\begin{aligned}
p\left(\mathcal{G}^{1:T}|\mathcal{G}^0\right)&=\prod_{t=1}^T p\left( \mathcal{G}^t|\mathcal{G}^{t-1}\right) = \prod_{t=1}^Tp\left((\mathcal{V}^{t}, \mathcal{E}^{t})|(\mathcal{V}^{t-1}, \mathcal{E}^{t-1})\right) \\
&=\prod_{t=1}^T (\mathcal{V}^0\cdot {Q}_\mathcal{V}^t,\mathcal{E}^0\cdot {Q}_\mathcal{E}^t) = (\mathcal{V}^0\cdot\tilde{Q}_\mathcal{V}^T,\mathcal{E}^0\cdot\tilde{Q}_\mathcal{E}^T),
\end{aligned}
\end{equation}
where $\tilde{Q}_\mathcal{V}^t = {Q}_\mathcal{V}^1 \cdots {Q}_\mathcal{V}^T$, $\tilde{Q}_\mathcal{E}^t = {Q}_\mathcal{E}^1 \cdots {Q}_\mathcal{E}^T$, and $\mathcal{G}^{1:T}$ denotes the sequence of intermediate graphs $\mathcal{G}^{1}, \mathcal{G}^{2}, \ldots,\mathcal{G}^{T}$.

\subsubsection{Graph Denoising Stage} The graph denoising stage serves as the inverse of the diffusion process, aiming to recover a clean graph $\mathcal{G}^{0}$ that maximizes the utility defined in (\ref{utility_function}) by inputting a noisy graph $\mathcal{G}^{t}=\left(\mathcal{V}, \mathcal{E}\right)$ into the denoising neural network $\phi_{\vartheta}$ parameterized by $\vartheta$. Thus, each denoising step generates a less noisy graph $\mathcal{G}^{t-1}$ from $\mathcal{G}^{t}$, and can be characterized by the conditional probability
\begin{equation}\label{stimate_the_reverse_diffusion_iterations}
q_\vartheta(\mathcal{G}^{t-1}|\mathcal{G}^t)=\prod_{1\leq i\leq n}q_\vartheta(v_i^{t-1}|\mathcal{G}^t)\prod_{1\leq i,j\leq n}q_\vartheta(e_{ij}^{t-1}|\mathcal{G}^t).
\end{equation}
To compute each term in (\ref{stimate_the_reverse_diffusion_iterations}), we marginalize over the neural network predictions. For the set of nodes, this yields 
\begin{equation}\label{network_predictions}
\begin{aligned}
q_\vartheta(v_i^{t-1}|\mathcal{G}^t)&=\int_{v_i}q_\vartheta(v_i^{t-1}\mid v_i^{0},\mathcal{G}^t)d q_\vartheta(v_i|\mathcal{G}^t)\\
    &=\sum_{v\in\mathcal{V}}q_\vartheta(v_i^{t-1}\mid v_i=v,\mathcal{G}^t)\bar{q}_i^V(v),
\end{aligned}
\end{equation}
where $q_\vartheta(v_i^{t-1}|v_i^0=v,\mathcal{G}^t)=p(v_i^{t-1}|v_i^0=v,v_i^t)$, the posterior distribution $\bar{q}_i^V(v)$ is a prediction based on $v \in \left\{0, 1\right\}$ given a noisy graph $\mathcal{G}^t$.
Similarly, for the set of edges, we obtain
\begin{equation}
\begin{aligned}q_\vartheta(e_{ij}^{t-1}|e_{ij}^t)&=\sum_{e\in\mathcal{E}}q_{\vartheta}(e_{ij}^{t-1}|e_{ij}^{0}=e,\mathcal{G}^{t})\bar{q}_{ij}^{E}(e|\mathcal{G}^{t}),\\
    &\begin{aligned}=\sum_{e\in\mathcal{E}}p(e_{ij}^{t-1}|e_{ij}^0=e,e_{ij}^t)\bar{q}_{ij}^E(e|\mathcal{G}^t).\end{aligned}\end{aligned}
\end{equation}
In summary, these distributions allow us to sample a discrete $\mathcal{G}^{t-1}$ as the input of the denoising network at the next time step, and each denoising step is formulated as
\begin{equation}\label{denoising_step}
q_\vartheta(\mathcal{G}^{t-1}|\mathcal{G}^t)=\sum_{\mathcal{G}\in \Omega}p(\mathcal{G}^{t-1}|\mathcal{G}^0=\mathcal{G},\mathcal{G}^t)\bar{q}_\vartheta(\mathcal{G}|\mathcal{G}^t),
\end{equation}
where $\Omega$ denotes the set of all possible initial graphs, and $\bar{q}_\vartheta(\mathcal{G}|\mathcal{G}^t)$ is the output distribution of denoising network. From (\ref{denoising_step}), we can express the denoising stage as a $T$-step Markov decision process (MDP) that can be solved by the RL algorithm. Simarly to~\cite{liu2024graph}, the MDP is formulated as
\begin{equation}
\begin{aligned}
\ \left\{\begin{array}{ll}
\boldsymbol{s}_{t}\triangleq(\mathcal{G}_{T-t},T-t),\boldsymbol{a}_{t}\triangleq\mathcal{G}_{T-t-1}, \\
\pi_\vartheta(\boldsymbol{a}_t|\boldsymbol{s}_t)\triangleq q_\vartheta(\mathcal{G}_{T-t-1}|\mathcal{G}_{T-t}), \\
r(\boldsymbol{s}_{t},\boldsymbol{a}_{t})\triangleq r(\mathcal{G}^{0}), \ \mathrm{if} \ t=T,\\
r(\boldsymbol{s}_{t},\boldsymbol{a}_{t})\triangleq0, \ \mathrm{if} \ t<T,
\end{array}\right.
\end{aligned}
\end{equation}
where $\boldsymbol{s}_{t}$ and $\boldsymbol{a}_{t}$ denote the sate and action at the $t$-step respectively, $\pi_\vartheta$ represents the policy of joint AeBS deployment and user association, $r(\boldsymbol{s}_{t},\boldsymbol{a}_{t})$ denotes the reward signal, $\mathcal{G}^{0}$ denotes the generated graph state after $T$ denoising steps. The sequence of graphs $\left(\mathcal{G}^{T}, \mathcal{G}^{T-1}, \ldots, \mathcal{G}^{0}\right)$ forms a trajectory $z$ of state-action pairs in this MDP. Furthermore, the accumulative reward is given by
\begin{equation}
R(z)=\sum_{t=0}^{T}r(s_{t},\boldsymbol{a}_{t})=r(\mathcal{G}^{0}).
\end{equation}
Therefore, solving optimization sub-problem $\mathcal{P}2$ is transformed into maximizing the agent's expected cumulative reward which is shown as
\begin{equation}
\mathcal{R}(\vartheta)=\mathbb{E}_{q(z|\pi_{\vartheta})}[R(z)]=\mathbb{E}_{q_{\vartheta}(\mathcal{G}^{0:T})}[r(\mathcal{G}^{0})].
\end{equation}
Our goal is to adjust the denoising network parameter $\vartheta$ to maximize $\mathcal{R}(\vartheta)$, thereby training the proposed JSGD algorithm to produce the optimal deployment and association strategy. Effective policy gradient methods, such as RL and proximal policy optimization (PPO), have already been used to generate a policy trajectory~\cite{sutton2018reinforcement, schulman2017proximal}. Thus, the expected cumulative reward $\mathcal{R}(\vartheta)$ can be optimized via policy gradient, derived as~\cite{liu2024graph}
\begin{equation}\label{policy_gradient}
\nabla_\vartheta\mathcal{R}(\vartheta)=\mathbb{E}_z\left[r(\mathcal{G}^0)\sum_{t=0}^{T-1}\nabla_\vartheta\log q_\vartheta(\mathcal{G}^0|\mathcal{G}^t)\right],
\end{equation}
where the expectation is taken over trajectories $z$ following $q_{\vartheta}(G_{0:T})$. Since the exact policy gradient $\nabla_\vartheta\mathcal{R}(\vartheta)$ in (\ref{policy_gradient}) is generally intractable to compute, we employ Monte Carlo sampling to approximate it:
\begin{equation}\label{discrete_policy_gradient}
\nabla_\vartheta\mathcal{R}(\vartheta)\approx\frac{1}{M}\sum_{m=1}^{M}\frac{T}{|\mathcal{T}_{m}|}\sum_{t\in\mathcal{T}_{m}}r(\mathcal{G}_{m}^{0})\nabla_{\vartheta}\mathrm{log}q_{\vartheta}(\mathcal{G}_{m}^{t-1}|\mathcal{G}_{m}^{t}),
\end{equation}
where $\{\mathcal{G}_{m}^{0:T}\}_{m=1}^{M}$ are $M$ trajectories sampled from $q_{\vartheta}(\mathcal{G}^{0:T})$ and $\{\mathcal{T}_{m}\subset[1,T]\}_{m=1}^{M}$ are random subsets of timesteps. Furthermore, (\ref{discrete_policy_gradient}) can be changed as a eager policy gradient to improve the efficiency of exploring high reward\footnote{Due to the large scale of the RSMA-assisted LAWNs, the graph generated by the RL policy gradient is difficult to converge to high reward values. The eager policy gradient leverages the prior graph $\mathcal{G}^0$ to group graph trajectories into equivalence classes, which effectively reduces the trajectory search space and enhances the efficiency of identifying higher reward regions~\cite{liu2024graph}.}, which is shown as
\begin{equation}\label{discrete_eager_policy_gradient}
g(\vartheta)\approx\frac{1}{M}\sum_{m=1}^{M}\frac{T}{|\mathcal{T}_{m}|}\sum_{t\in\mathcal{T}_{m}}r(\mathcal{G}_{m}^{0})\nabla_{\vartheta}\mathrm{log}q_{\vartheta}(\mathcal{G}_{m}^{0}|\mathcal{G}_{m}^{t}).
\end{equation}

\begin{figure*}
    \centering
    \includegraphics[width=0.99\linewidth]{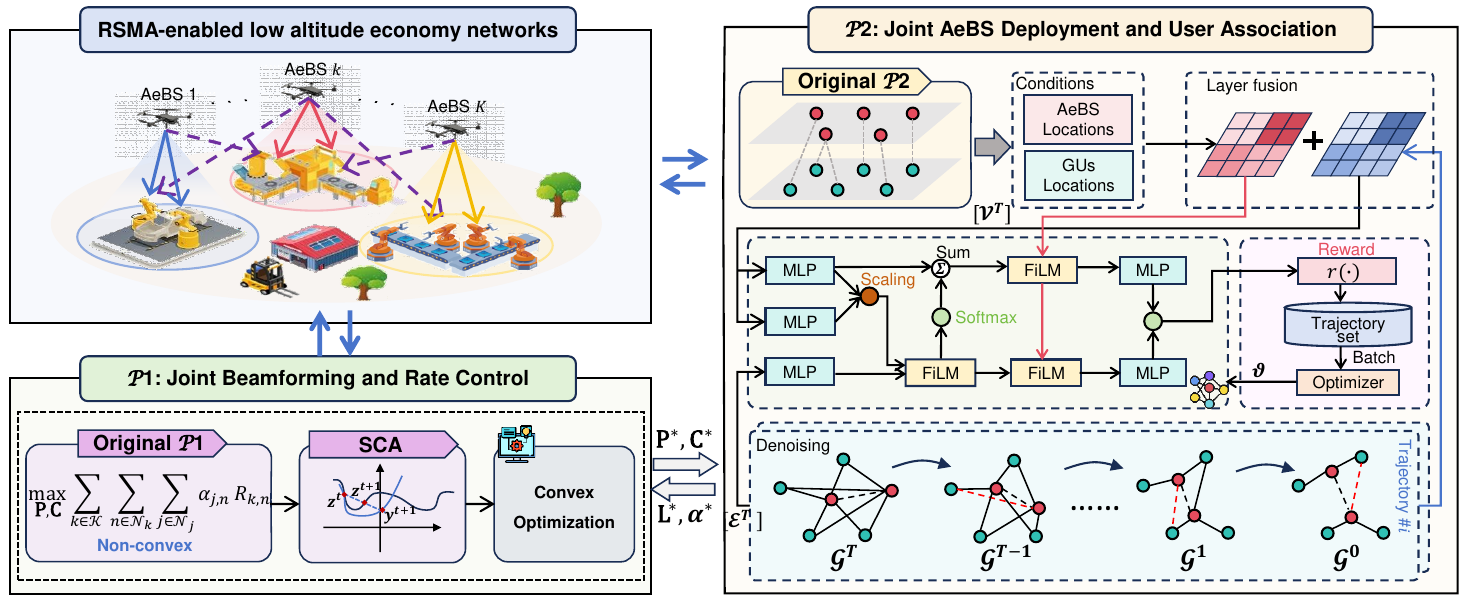}
    \caption{\small{The framework of the proposed JSGD algorithm for AeBS deployment, user association, and resource allocation in RSMA-enabled low-altitude wireless networks.}}
    \label{figure_algorithm_framework}
\end{figure*}

\subsubsection{Reward Function Design}To provide efficient feedback for each generated graph $\mathcal{G}^0$ and to guide the training of JSGD, we must carefully design the reward function $r\left(\cdot \right)$.
Considering the optimization objective and constraints in (\ref{subproblem_2}), a penalty-constrained reward function is designed as
\begin{equation}\label{reward}
r(\mathcal{G}_{m}^{0}) =\mathcal{U}  - \omega_{1}\xi_{a} - \omega_{2}\xi_{m} - \omega_{3}\xi_{c} - \omega_{4}\xi_{r},
\end{equation}
where the first term is a positive reward relevant to the system utility, and the latter four terms are the penalty functions corresponding to association constraints C1 $\sim$ C3, location constraints C4, collision constraint C9, and communication range constraint C10. Specifically, $w_{\ell}, \ell \in \left\{1, 2, 3, 4\right\}$ are the penalties applied if constraints are violated. $\xi_{a}$, $\xi_{m}$, $\xi_{c}$, and $\xi_{r}$ are binary indicators representing whether AeBS $k$ violates the user association constraints, mobility constraints, collision constraint, and communication range constraint, respectively.

\subsection{The Proposed JSGD Framework}

The proposed JSGD framework is shown in Fig.~\ref{figure_algorithm_framework}, where the environment continuously interacts with both the graph diffusion module and the SCA module. The graph diffusion module employs a graph-aware transformer architecture to perform denoising. In the layer fusion stage, we use one-hot encodings for the input AeBS/GU locations and user associations to process these embeddings and construct a randomly initialized graph $\mathcal{G}^T$. Within the denoising network, these embeddings are further mapped through multi-layer perceptrons (MLPs). The encoded embeddings are then fed into a feature-wise linear modulation (FiLM)-based graph transformer layer. This layer integrates cross-feature attention while adapting its output through learned affine transformations conditioned on the input features. 
During training, each denoising trajectory $\mathcal{G}_{m}^{0:T}$ is evaluated using the reward (\ref{reward}). The gradient of this utility is propagated backward through the denoising network, enabling end-to-end optimization of the generative policy.

\begin{algorithm}[!t]
    \caption{The Training Phase of the Proposed JSGD Algorithm for RSMA-enabled LAWNs}
    \label{training_algorithm}
    \SetKwData{Left}{left}\SetKwData{This}{this}\SetKwData{Up}{up}
 \SetKwFunction{FindCompress}{FindCompress}
	\SetKwInOut{Input}{Input}\SetKwInOut{Output}{Output}
	
	\Input{Initial denoising network parameter $q_\vartheta$\; a Markov decision process $\mathcal{M} = \left(\mathcal{S},\mathcal{A},q_{\vartheta}, r, \epsilon_{0}\right)$; denoising steps $T$; trajectory samples $M$; reward signal $r\left(\cdot\right)$; timestep samples $|\mathcal{T}|$; learning rate $\mu$; training steps $L$; $l \gets 0$; $i\gets0$; $t\gets0$.}

    \Repeat{$l \ge L$}{
    \Repeat{$i \ge M$}{
      Sample initial noisy graph from the MDP $\mathcal{M}$; \\ 
      According to the given position and user association, obtain the AeBS Beamforming $\mathbf{P}$ and common rate control $\mathbf{C}$ based on Algorithm~\ref{SCA_algorithm}; \\
      \Repeat{$t \ge T$}{
      Perform denoising based on (\ref{denoising_step});
      }
      Sample trajectory $\mathcal{G}^{0:T}_{i}$ from denoising network $q_\vartheta$; \\ 
      Sample timesteps ($\mathcal{T}_{m} \sim \text{Uniform}([1, T]$); \\
      For given $\mathbf{P}$ and $\mathbf{C}$, calculate rewards $r(\mathcal{G}_{m}^{0})$ according to (\ref{reward}); \\
      Update trajectory sample $i \gets i+1$;
    }
    Estimate the eager policy gradient $g(\vartheta)$ based on (\ref{discrete_eager_policy_gradient}); \\
    Update the model parameter $\vartheta \gets \vartheta + \mu \cdot g(\vartheta)$; \\
    Update training step $l \gets l+1$;
    }
\Output{Optimal denoising network. }
\end{algorithm}


The training phase of JSGD is shown in Algorithm~\ref{training_algorithm}. 
And the testing phase is shown in Algorithm~\ref{testing_algorithm}. 
In the testing phase, the trained graph diffusion model and SCA algorithm are alternately updated until convergence, yielding optimized AeBS deployment, user association, and resource allocation. Under the SCA framework, the diffusion model optimizes only AeBS positions and user associations rather than the full decision space, thereby reducing state–action dimensionality, especially in multi-AeBS coordination scenarios. 
Another novel aspect of the proposed approach lies in utilizing SCA to generate high-quality beamforming and common-rate allocations rather than discretizing power levels into a few predefined options as adopted in~\cite{8654727}.

\subsection{Complexity Analysis}

We denote the number of GUs associated with AeBSs in $l$-th episode as $N_{u,l}$. For the SCA-based resource allocation, since problem $\mathcal{P}1\textit{-}2$ involves $(5 + N_t)(K + N_{u,l})$ variables and $K + 8N_{u,l}$ constraints which are solved by the CVX solver within polynomial time, the complexity can be approximated as $\mathcal{O}(8N_t^3N_{u,l}^4)$. Let $W$ represent the number of graph transformer layers. In the GGDM-based AeBS deployment and user association, for $T$ diffusion steps, each step performs one-hot state transitions on $K$ nodes and $K \times N$ edges. Thus, the complexity of the graph diffusion stage is $O(T (K+KN))$, which can be approximated as $O(TKN)$. In the graph denoising stage, JSGD collects $M$ trajectories, each consisting of $T$ denoising steps. Therefore, the complexity of the graph denoising stage is approximately $O(M T d (K+N))$, where $d$ denotes the hidden layer dimension. Accordingly, the overall complexity of the GGDM-based AeBS deployment and user association component is calculated as $O(TKN+M T d (K+N))$. Given $L$ training steps, the total complexity of the optimization framework is computed as $\mathcal{O}{\left(TKNL + MTdL(K+N)+8{\textstyle \sum_{l=1}^{L}(N_t^3N_{u,l}^4)} \right)}$.

\begin{algorithm}[!t]
    \caption{The Testing Phase of the Proposed JSGD Algorithm for RSMA-enabled LAWNs}
    \label{testing_algorithm}
    \SetKwData{Left}{left}\SetKwData{This}{this}\SetKwData{Up}{up}
 \SetKwFunction{FindCompress}{FindCompress}
	\SetKwInOut{Input}{Initialization}\SetKwInOut{Output}{Output}
	
	\Input{$k\gets 0$, $\mathbf{P}^{\left[k\right]}$, $\mathbf{C}^{\left[k\right]}$, $\mathbf{L}^{\left[k\right]}$ and $\boldsymbol{\alpha}^{\left[k\right]}$ to a~feasible solution, the convergence tolerance $\tau=10^{-3}$, maximum number of iterations $k_{max}$.}

    \Repeat{$k \ge k_{max}$ or $\left | \mathcal{U}^{\left[k\right]}-\mathcal{U}^{\left[k-1\right]} \right | < \tau$}{
    \Repeat{$t \ge T$}{
    Given $\mathbf{P}^{\left[k\right]}$ and $\mathbf{C}^{\left[k\right]}$, perform denoising process based on (\ref{denoising_step}), then we obtain $\mathbf{L}^{\left[k+1\right]}$ and $\boldsymbol{\alpha}^{\left[k+1\right]}$;
    }
      Update $\mathbf{P}^{\left[k+1\right]}$ and $\mathbf{C}^{\left[k+1\right]}$ with given $\mathbf{L}^{\left[k\right]}$ and $\boldsymbol{\alpha}^{\left[k\right]}$ using Algorithm \ref{SCA_algorithm}; \\
      $k \gets k+1$;
    }
\Output{ the solution $\boldsymbol{\alpha}^{*}$, $\mathbf{L}^{*}$, $\mathbf{P}^{*}$, and $\mathbf{C}^{*}$.}
\end{algorithm}

\section{Simulation Results and Discussion}\label{sec_simulation}

In this section, we implement the proposed JSGD method for RSMA-enabled LAWNs, and conduct extensive experiments to evaluate the proposed framework.

\subsection{Experimental Setup}

\subsubsection{Experimental Configuration} The experiments are conducted using Torch 2.5.1 and Python 3.12.8, and the network training is implemented on a server running Ubuntu 22.04, equipped with an NVIDIA A100 GPU (80 GB), a 2.10 GHz Intel Xeon Gold 5218R processor with 40 cores, and 503 GB of RAM. The simulation scenario considers a $1000~\text{m} \times 1000~\text{m}$ task region, within which the AeBSs are dispatched to serve the GUs which are randomly distributed. Unless otherwise stated, the optimization preference factors are set to $\lambda_{1} = 1$ and $\lambda_{2} = 1$ respectively. Additionally, we assume that the minimum rate requirement of each GU is the same, i.e., $R_{\mathrm{min}} = R^{th}/N$. The main simulation parameters are presented in Table~\ref{table_parameters}.

\begin{table}[!t]\footnotesize
	\renewcommand{\arraystretch}{1.3}
	\caption{System Parameters}
	\label{table_parameters}
	\centering
	\begin{tabular}{p{0.65\linewidth}|p{0.25\linewidth}}
		\hline\hline
		\textbf{Parameters}
		& \textbf{Values}
		\\ \hline
            Number of transmitting antennas $N_t$
		& $4$
            \\ \hline
            LoS and NLoS attenuation factors $\zeta_{\mathrm{LoS}}$, $\zeta_{\mathrm{NLoS}}$
		& $1~\mathrm{dB}$, $20~\mathrm{dB}$~\cite{7510820}
		\\ \hline
            Speed of light $c$
		& $3 \times 10^{8}~\mathrm{m/s}$
		\\ \hline
            Environmental parameters $\eta$, $\varsigma$
		& $9.61$, $0.16$~\cite{7510820}
            \\ \hline
            Carrier frequency $f_c$ & $2.4~\mathrm{GHz}$
            \\ \hline
            Maximum transmission power $P_{\mathrm{max}}$ & $10~\mathrm{dBm}$
            \\ \hline
            Power of noise & $-113~\mathrm{dBm}$
            \\ \hline
            Safe distance between AeBSs $d_{\mathrm{min}}$ & $10~\mathrm{m}$~\cite{9963915}
            \\ \hline
            Flight hight $H$ & $50~\mathrm{m}$
            \\ \hline
            The communication radius $R$ & $100 \sim 250~\mathrm{m}$
            \\ \hline
            Number of AeBSs $K$ & $2 \sim 5$
            \\ \hline
            Number of GUs $N$ & $10 \sim 18$
            \\ \hline
            Channel blocklength $D^{c}$, $D^{p}$ & $1000$
            \\ \hline
            Latency budget $T_{th}$ & $1~\mathrm{ms}$
            \\ \hline
            Available bandwidth $B$ & $2~\mathrm{MHz}$
            \\ \hline
            Decoding error probability $\varepsilon$ & $10^{-5}$
		\\ \hline
            Maximum GU capacity of AeBS $N_a$ & $10$
		\\ \hline
            Minimal rate requirement $R_{\mathrm{min}}$ & $1~\mathrm{bit/s/Hz}$
		\\ \hline
		Convergence tolerance $\tau$
		& $10^{-5}$
            \\ \hline
		  The number of denoising steps $T$
		& $5$, $15$, $25$
            \\ \hline
		  Learning rate $\mu$
		& $10^{-5}$
            \\ \hline
		  Batch size & $64$
		\\ \hline\hline
	\end{tabular}
\end{table}

\subsubsection{Baseline Settings} To present fair comparison, the performance of the proposed JSGD algorithm for RSMA-enabled LAWNs (\textbf{JSGD-RS}) is evaluated versus the following baselines: joint SCA and DRL optimization for RSMA-enabled LAWNs (\textbf{JSDR-RS}), where AeBSs adopt a conventional policy-based DRL framework~\cite{10309220}, instead of generative models, to obtain the location and user association results, and adopt the beamforming and common rate control method based on Algorithm~\ref{SCA_algorithm} in RSMA-enabled LAWNs; the proposed JSGD optimization for NOMA-enabled LAWNs (\textbf{JSGD-NO}), where AeBSs adopt the proposed JSGD algorithm to obtain the location, user association and beamforming results in NOMA-enabled LAWNs; the proposed JSGD optimization for SDMA-enabled LAWNs (\textbf{JSGD-SD}), where AeBSs adopt the JSGD algorithm to obtain the location, user association and beamforming results in SDMA-enabled LAWNs; and \textbf{Random}, where the AeBSs randomly choose the location, user association and resource allocation in RSMA-enabled LAWNs. Note that, to ensure fairness, the JSDR baseline is set to the same setting as JSGD algorithm.

\subsection{Experimental Results}

\begin{figure}[!t]
\centering
\includegraphics[width=0.8\linewidth]{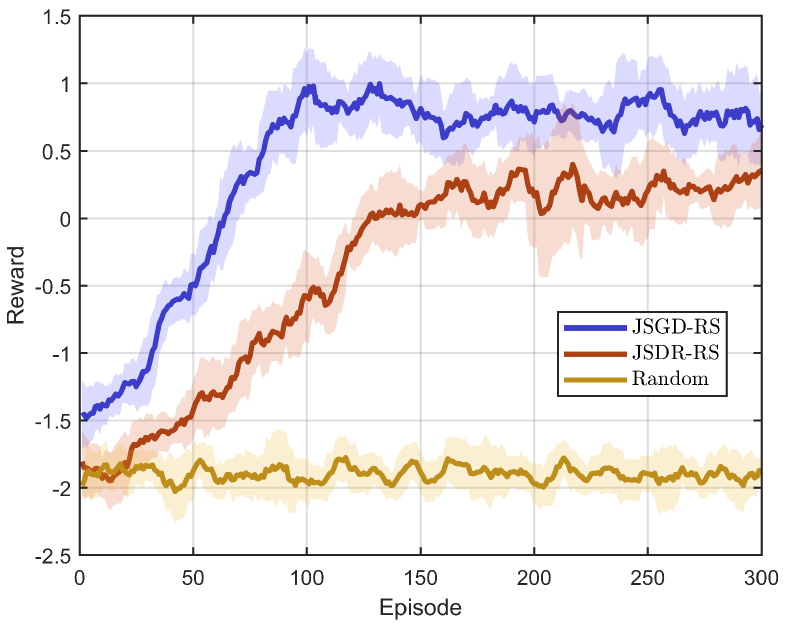}
\caption{The reward curves of the proposed JSGD algorithm and other baselines for RSMA-enabled LAWNs over training episodes.}
\label{fig:convergence}
\end{figure}

Fig.~\ref{fig:convergence} illustrates the reward curves of the proposed JSGD and other baselines for RSMA-enabled LAWNs. We can observe that the proposed JSGD-RS algorithm achieves a notably higher reward value, stabilizing around $0.7493$, whereas the JSDR-RS baseline only reaches approximately $0.1235$. The performance gap indicates that the graph diffusion architecture can better exploits environmental structure and interdependencies among UAV locations and user associations. 
Besides, the proposed JSGD-RS demonstrates faster convergence, stabilizing after approximately $100$ episodes, while the JSDR-RS method exhibits slower convergence with higher oscillations throughout the training process. This is because the diffusion mechanism naturally propagates information gradually between nodes, thereby enabling efficient representation learning. A high reward indicates that the generated optimization strategy can provide services with a high sum rate and high coverage, demonstrating the proposed JSGD is effective in optimizing communication service provisioning in LAWNs.

\begin{figure}[t!]
\centering
\subfloat[]{%
    \includegraphics[width=0.352\textwidth]{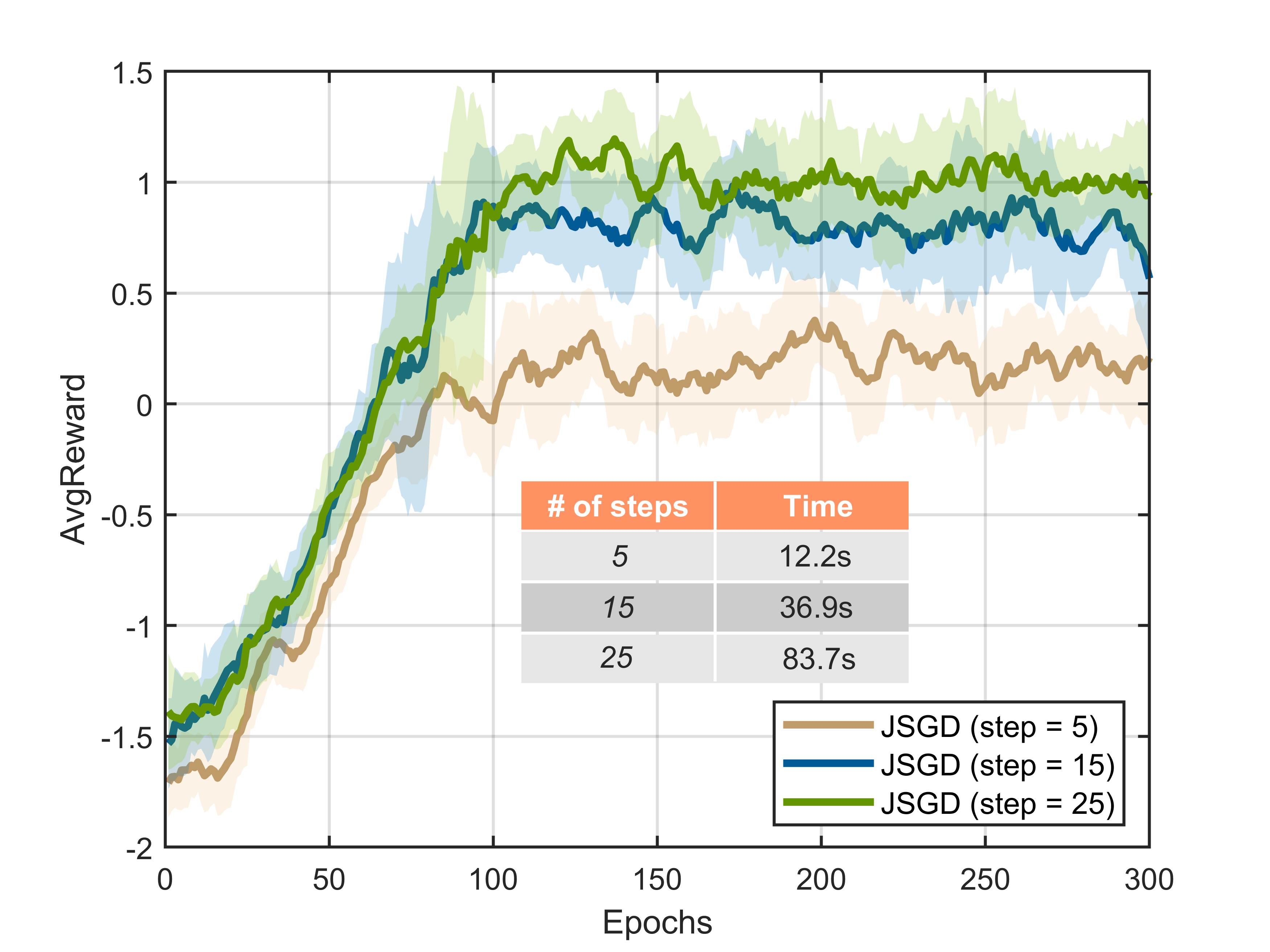}%
    \label{fig:convergence_vs_denoising_step}
}
\hspace{-7.3mm}
\subfloat[]{%
    \includegraphics[width=0.1565\textwidth]{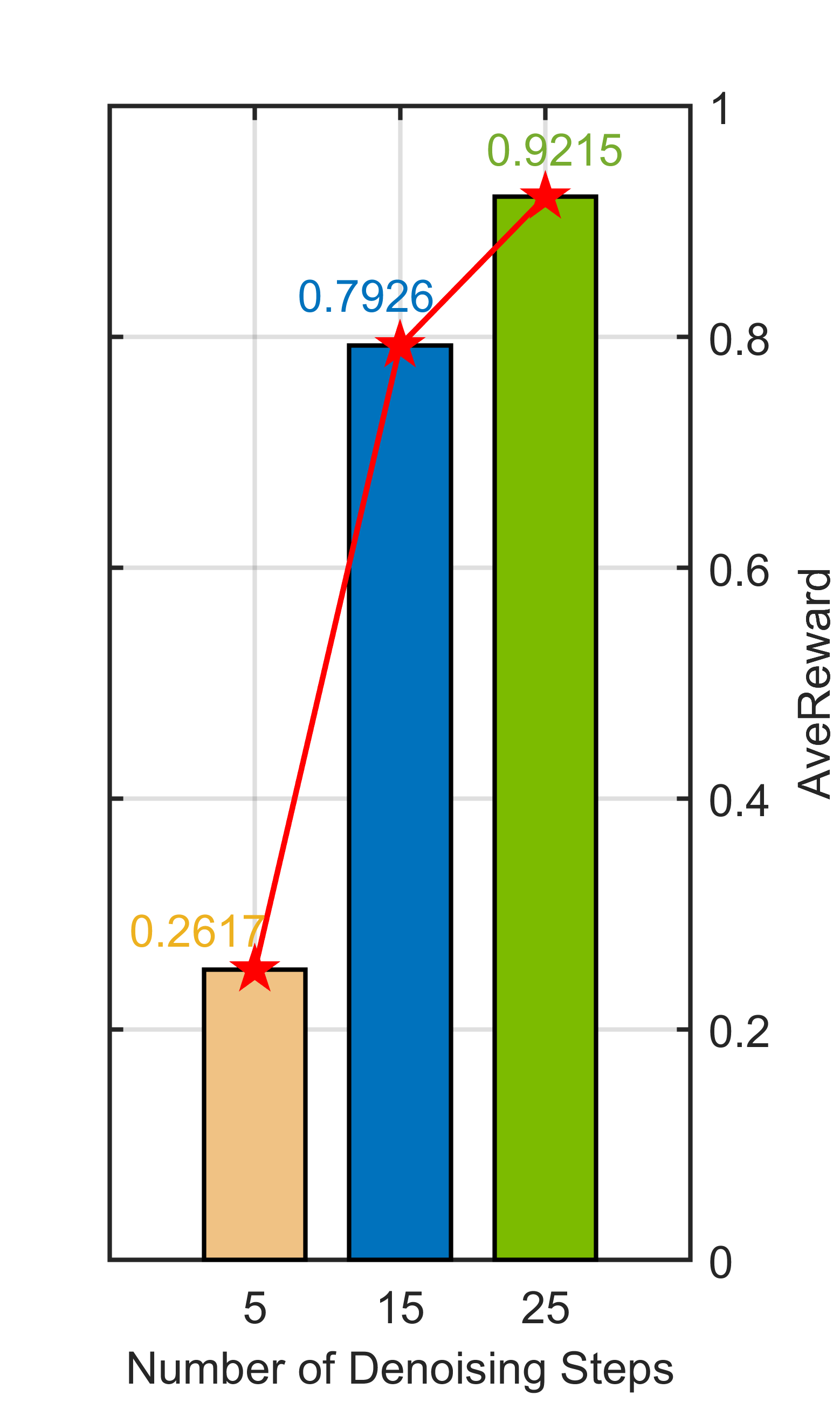}%
    \label{fig:convergence_vs_denoising_step_ZhuZhuangTu}
}
\caption{The reward curves of the proposed JSGD algorithm under different number of denoising steps.}
\label{fig:denoising_step}
\end{figure}

\begin{figure}[t!]
\centering
\subfloat[]{%
    \includegraphics[width=0.352\textwidth]{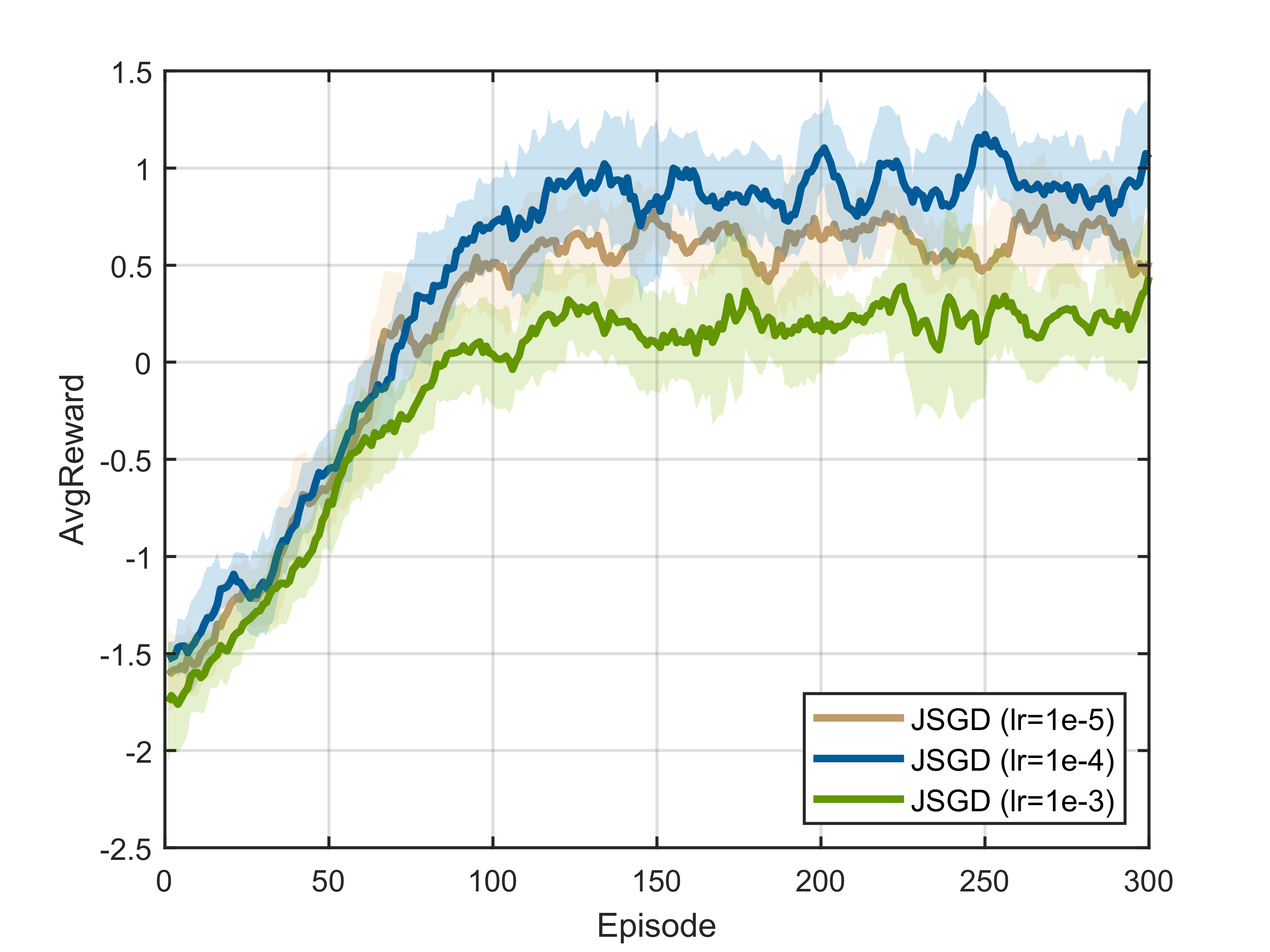}%
    \label{fig:convergence_vs_lr}
}
\hspace{-7.3mm}
\subfloat[]{%
    \includegraphics[width=0.1565\textwidth]{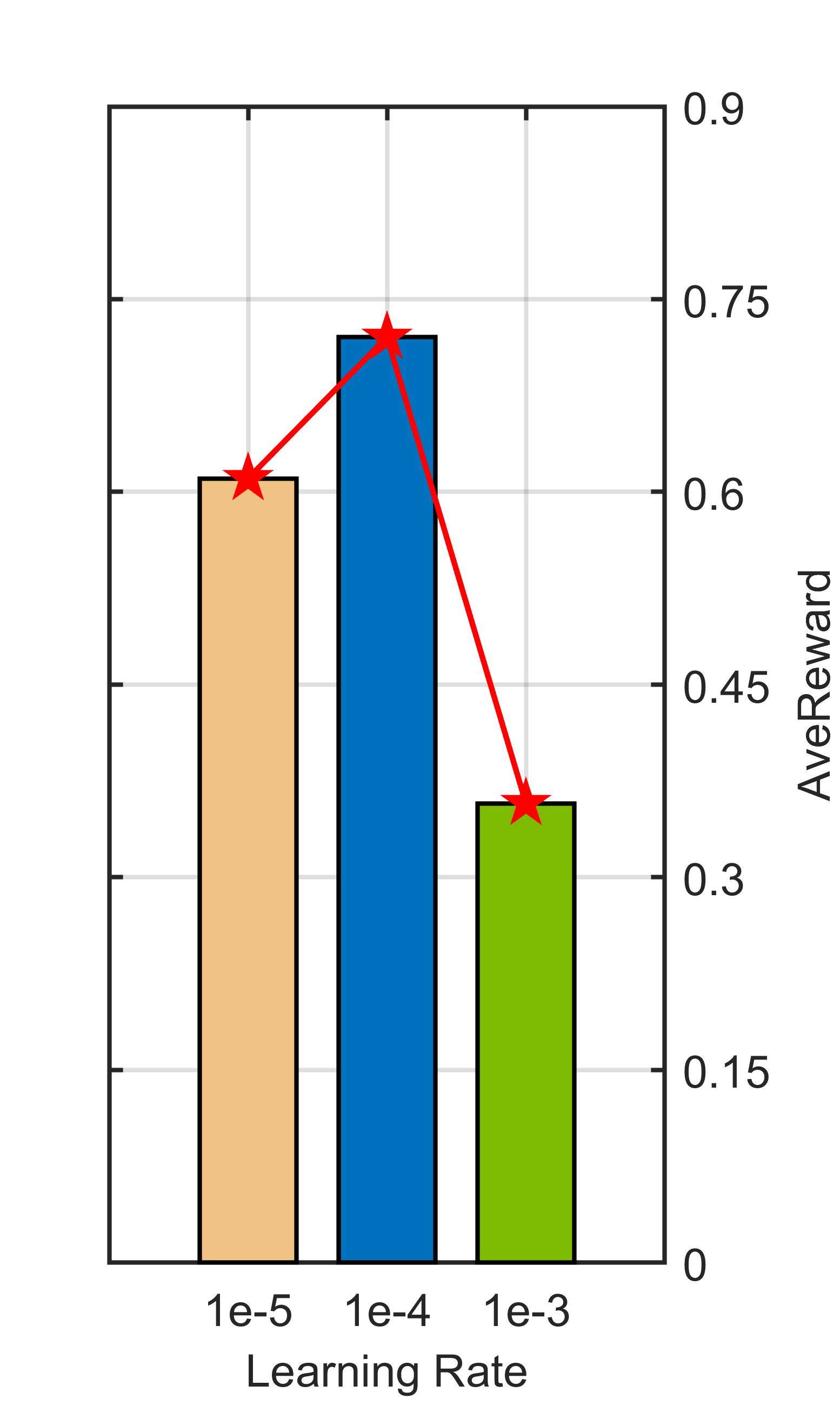}%
    \label{fig:convergence_vs_lr_ZhuZhuangTu}
}
\caption{The reward curves of the proposed JSGD algorithm under different learning rate.}
\label{fig:lr}
\end{figure}

Fig.~\ref{fig:denoising_step} illustrates the reward curves of the proposed JSGD algorithm under varying numbers of denoising steps. First, we observe that the average reward improves as the number of denoising steps increases, and the JSGD algorithm with $25$ denoising steps achieves the highest reward values, while the JSGD algorithm with $5$ denoising steps achieves the weakest performance due to insufficient noise reduction. This is because a greater number of denoising steps makes the noise-added distribution more closely approximate a standard normal distribution, which facilitates learning an accurate posterior distribution. However, increasing the number of denoising steps also results in higher computational time. In particular, as the number of denoising steps grows from $15$ to $25$, the reward improvement becomes insignificant, while the computation time noticeably increases from $36.9s$ to $83.7s$. Therefore we select $15$ denoising steps as a compromise, thereby achieving a balance between reward and training costs.

Fig.~\ref{fig:lr} depicts the performance of the proposed JSGD algorithm under various learning settings. Selecting a learning rate of $10^{-4}$ yields an average reward of 0.7205. The learning rate of $10^{-3}$ results in faster initial convergence; however, the final reward performance is significantly lower, indicating that excessive step sizes lead to unstable learning and degraded performance. In addition, the learning rate of $10^{-5}$ achieves lower reward performance than that of $10^{-4}$. This slower convergence is due to insufficient exploration, leading to the model becoming trapped in local optima. Therefore, the learning rate is set to $10^{-4}$ for all subsequent experiments.

\begin{figure*}
    \centering
    \includegraphics[width=0.95\linewidth]{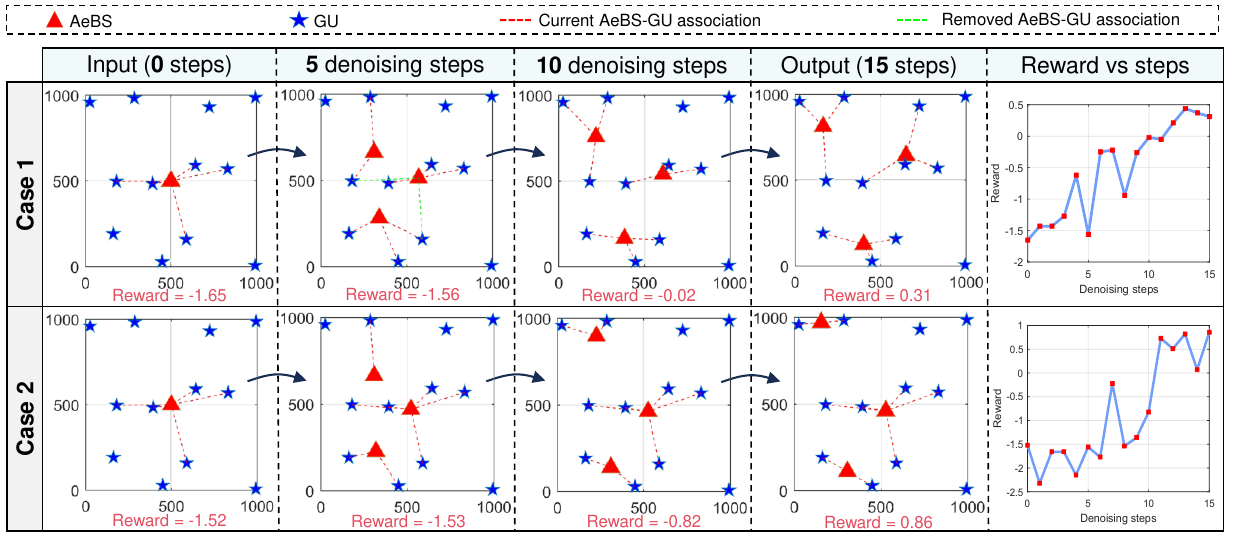}
    \caption{The graph generation process in RSMA-enabled LAWNs under two different hyperparameter settings, where the number of denoising steps $T=15$, the number of AeBSs $K=3$, and the number of GUs $N=12$. Case 1: $\lambda_{1} : \lambda_{2} = 2:1$. Case 2: $\lambda_{1} : \lambda_{2} = 1:2$.}
    \label{figure_deployment_vs_denoising_steps}
\end{figure*}

\begin{figure*}[!t]  
\centering
\subfloat[\tiny]{
    \includegraphics[width=0.32\linewidth]{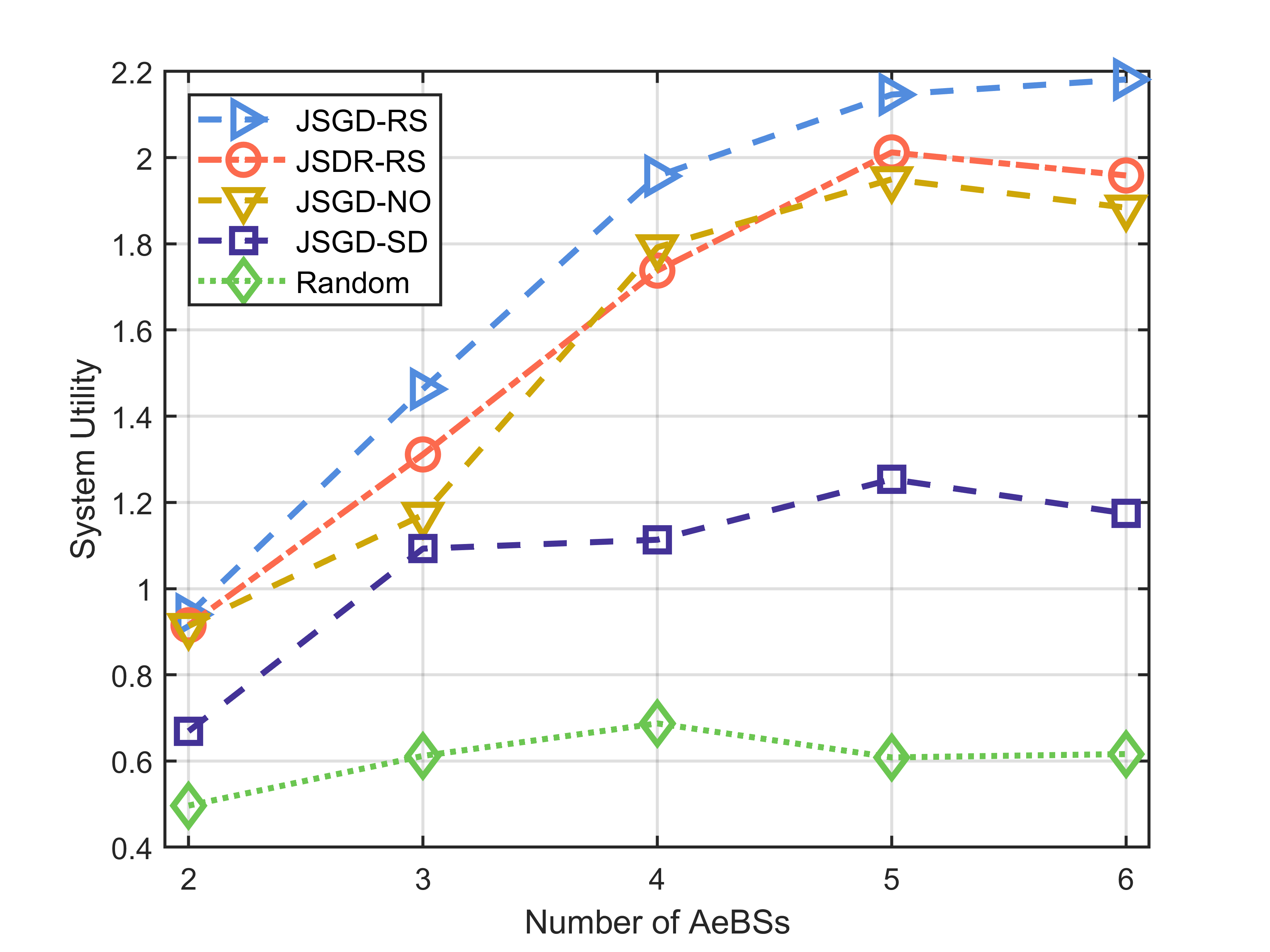} 
    \label{system_utility_vs_AeBS}
}
\subfloat[\tiny]{
    \includegraphics[width=0.32\linewidth]{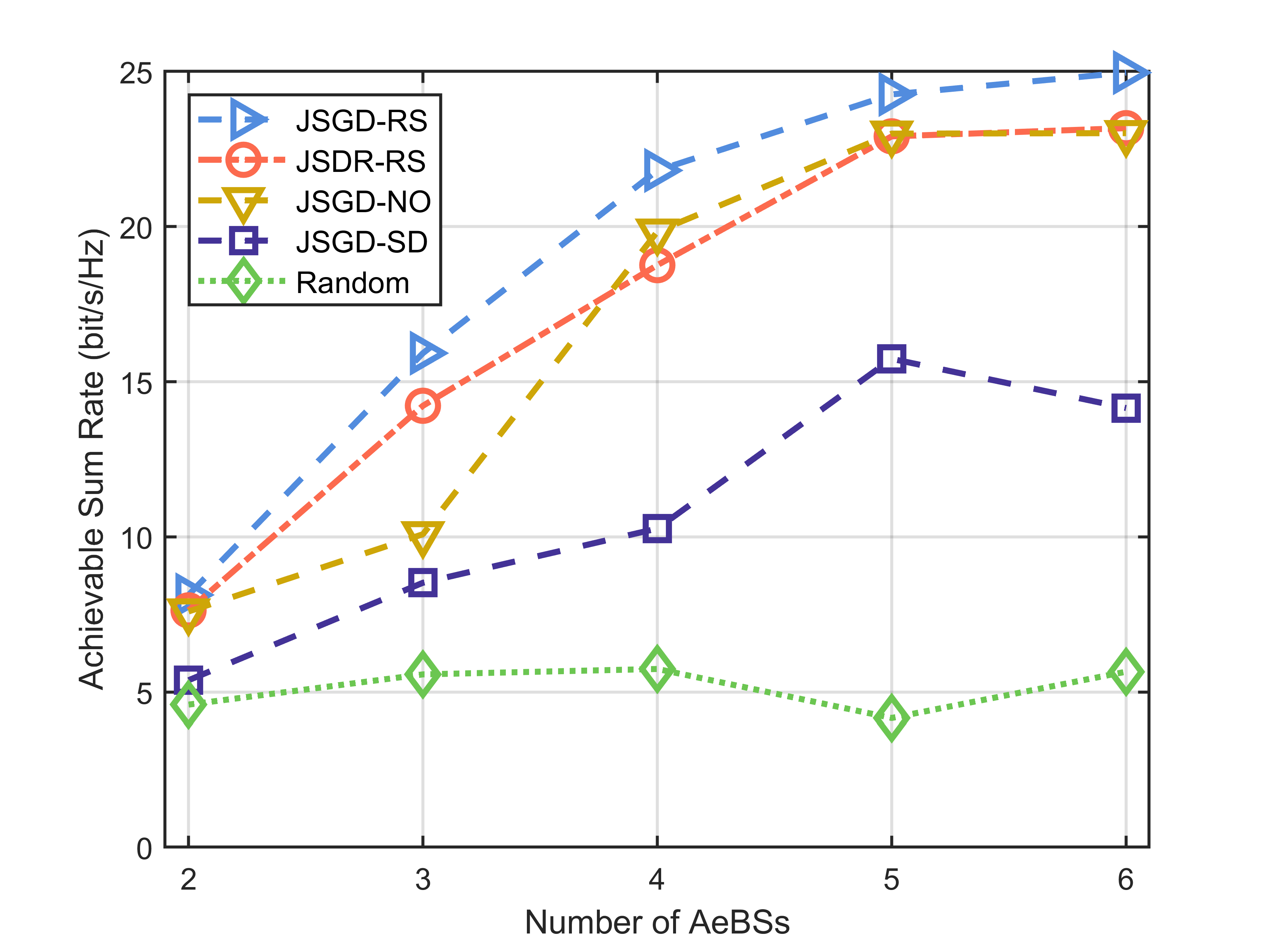}
    \label{sum_rate_vs_AeBS}
}
\subfloat[\scriptsize]{
    \includegraphics[width=0.32\linewidth]{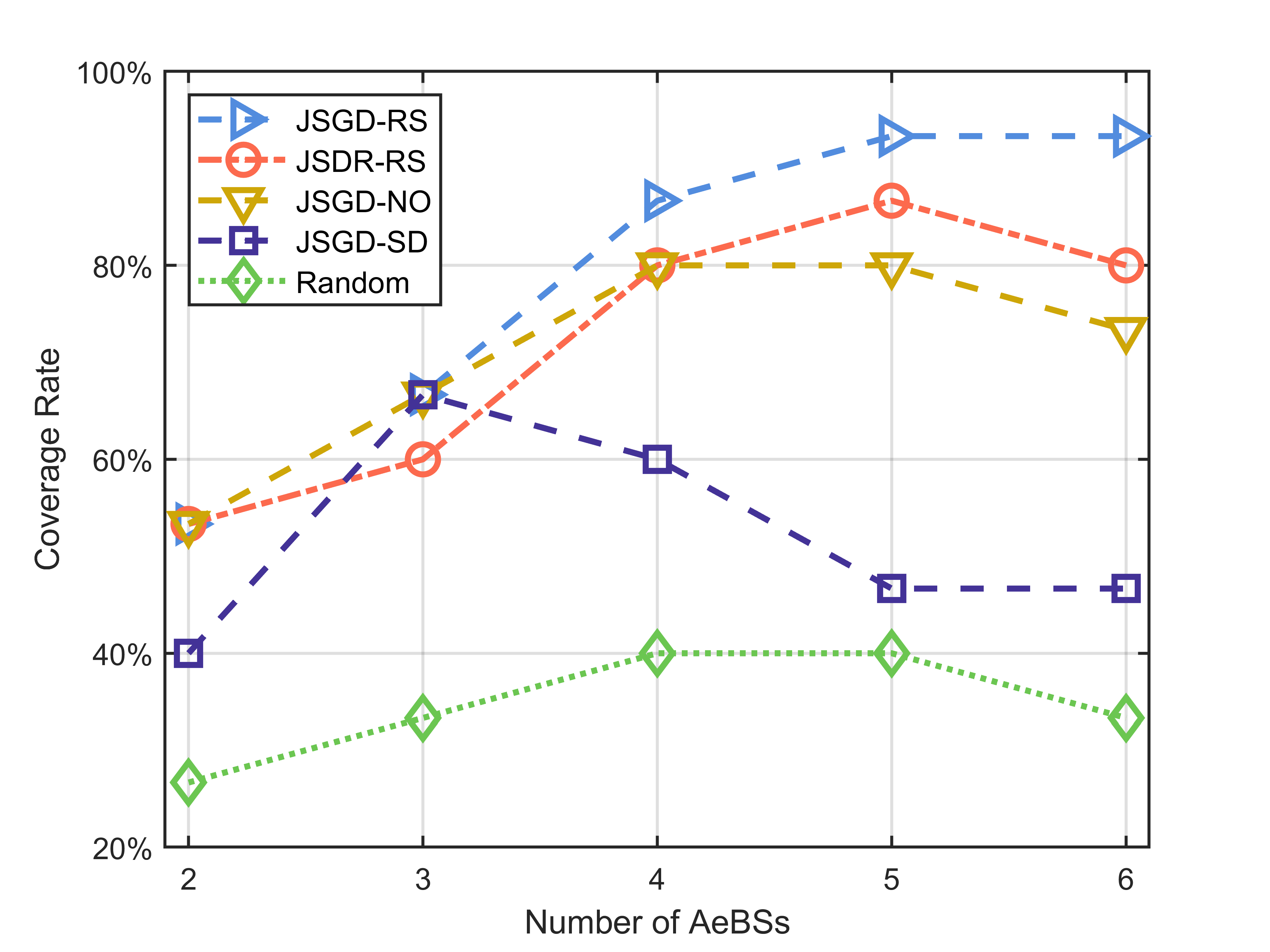}
    \label{fig_3}
}
\caption{Impact of the number of AeBSs on: (a) system utility, (b) achievable sum rate, (c) coverage rate.}
\label{impact_of_number_of_AeBS}
\end{figure*}

Fig.~\ref{figure_deployment_vs_denoising_steps} illustrates the graph generation process for RSMA-enabled LAWNs using our proposed JSGD framework under two different network cases, in which the initial positions of the AeBSs are set at $\left(500, 500\right)$. The displayed results demonstrate that the JSGD algorithm can intelligently adjust suboptimal node positions and edges to iteratively optimize the graph structure. Guided by the designed reward function, the graphs evolve towards optimized configurations, with the overall reward steadily increasing as the number of denoising steps grows, indicating the increasing quality of the generated graphs. These results confirm the effectiveness of our framework in learning to generate high-quality graphs under diverse system requirements, validating both the training of the denoising network and the design of the reward function.

\begin{figure*}[!t]  
\centering
\subfloat[]{
    \includegraphics[width=0.32\linewidth] {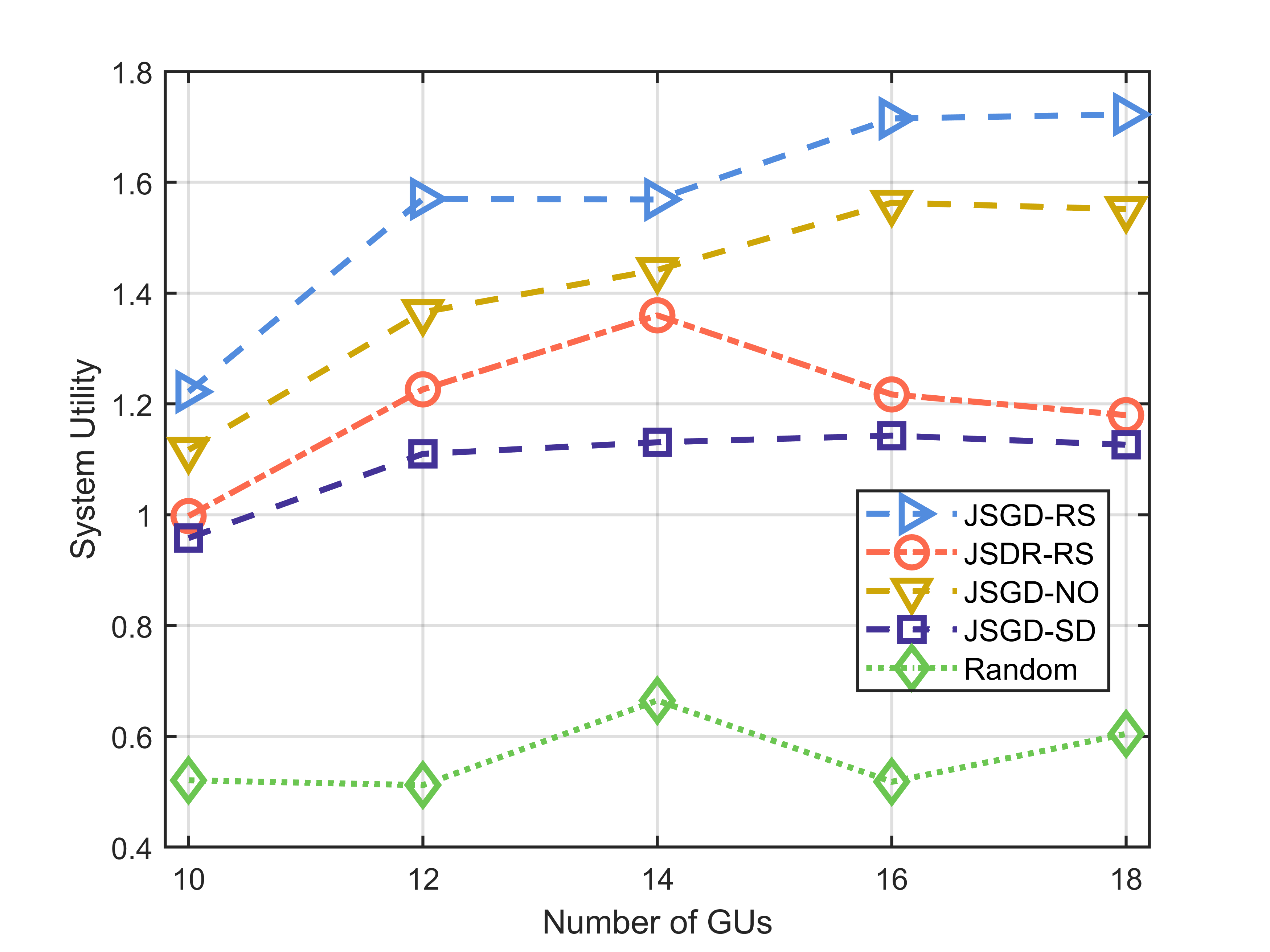} 
    \label{system_utility_vs_GU}
}
\subfloat[]{
    \includegraphics[width=0.32\linewidth]{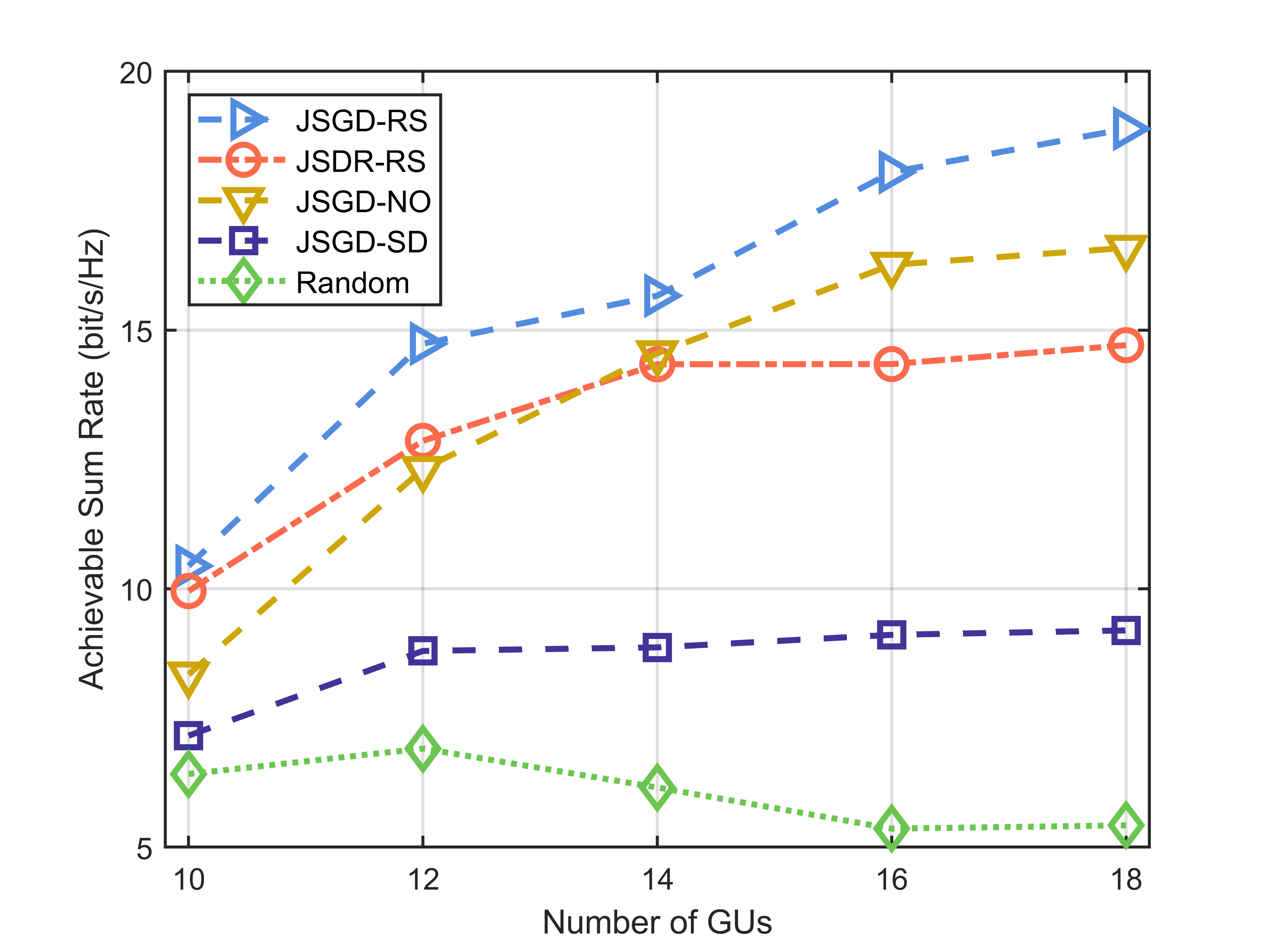}
    \label{sum_rate_vs_GU}
}
\subfloat[]{
    \includegraphics[width=0.32\linewidth]{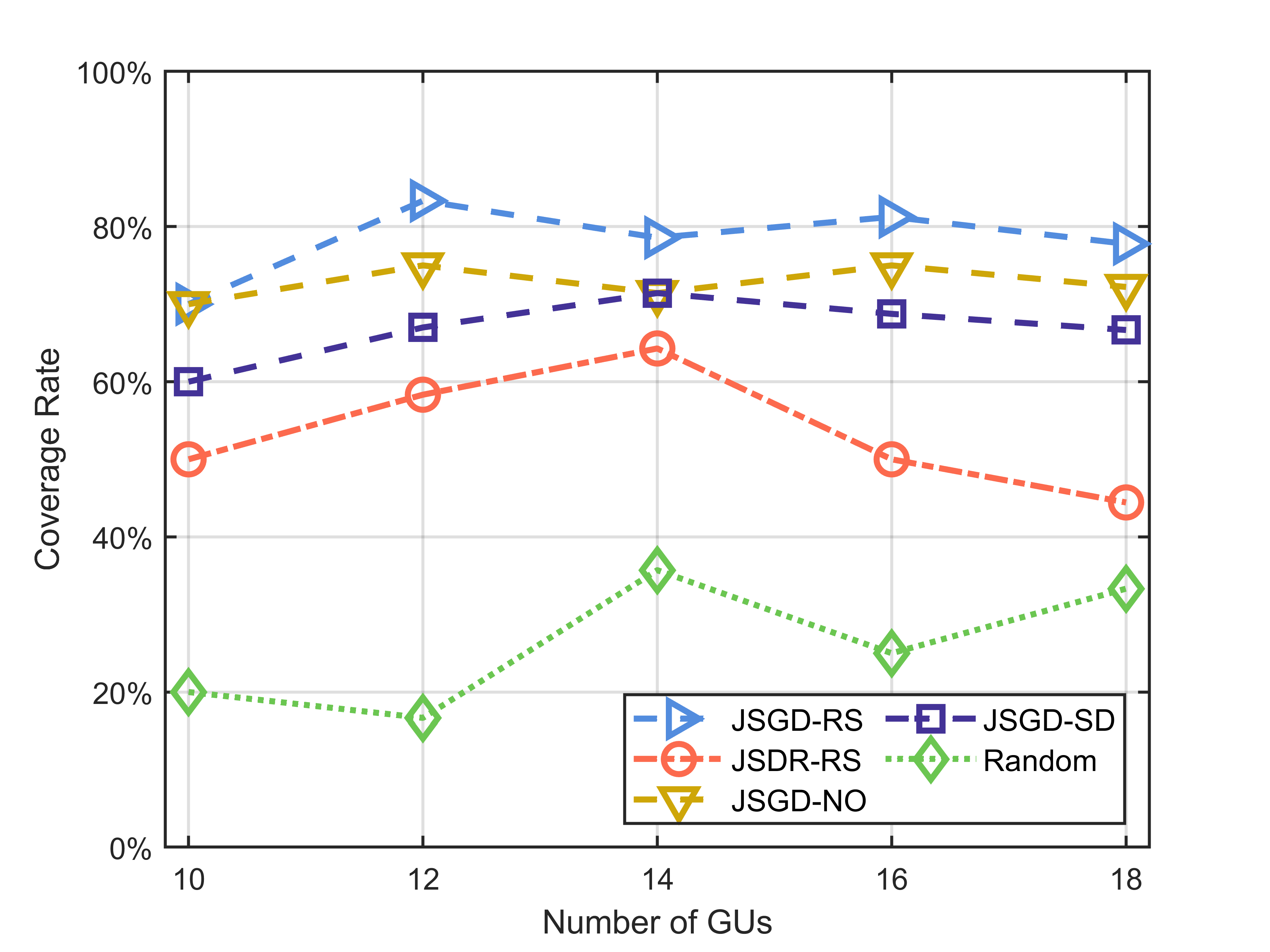}
    \label{coverage_vs_GU}
}
\caption{Impact of the number of GUs on: (a) system utility, (b) achievable sum rate, (c) coverage rate.}
\label{impact_of_number_of_GU}
\end{figure*}

\begin{figure*}[!t]  
\centering
\subfloat[]{
    \includegraphics[width=0.32\linewidth]{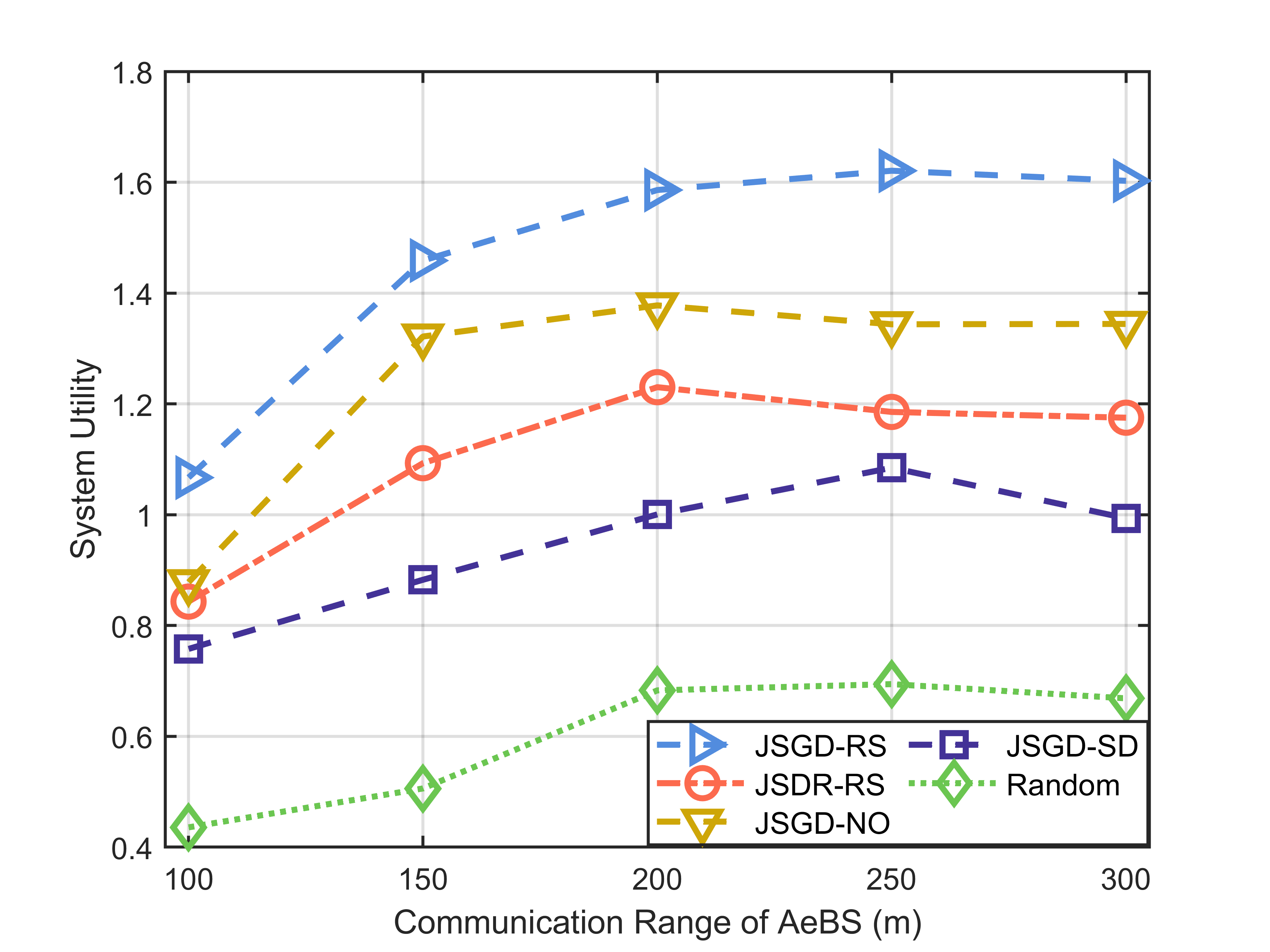} 
    \label{system_utility_vs_communication_range}
}
\subfloat[]{
    \includegraphics[width=0.32\linewidth]{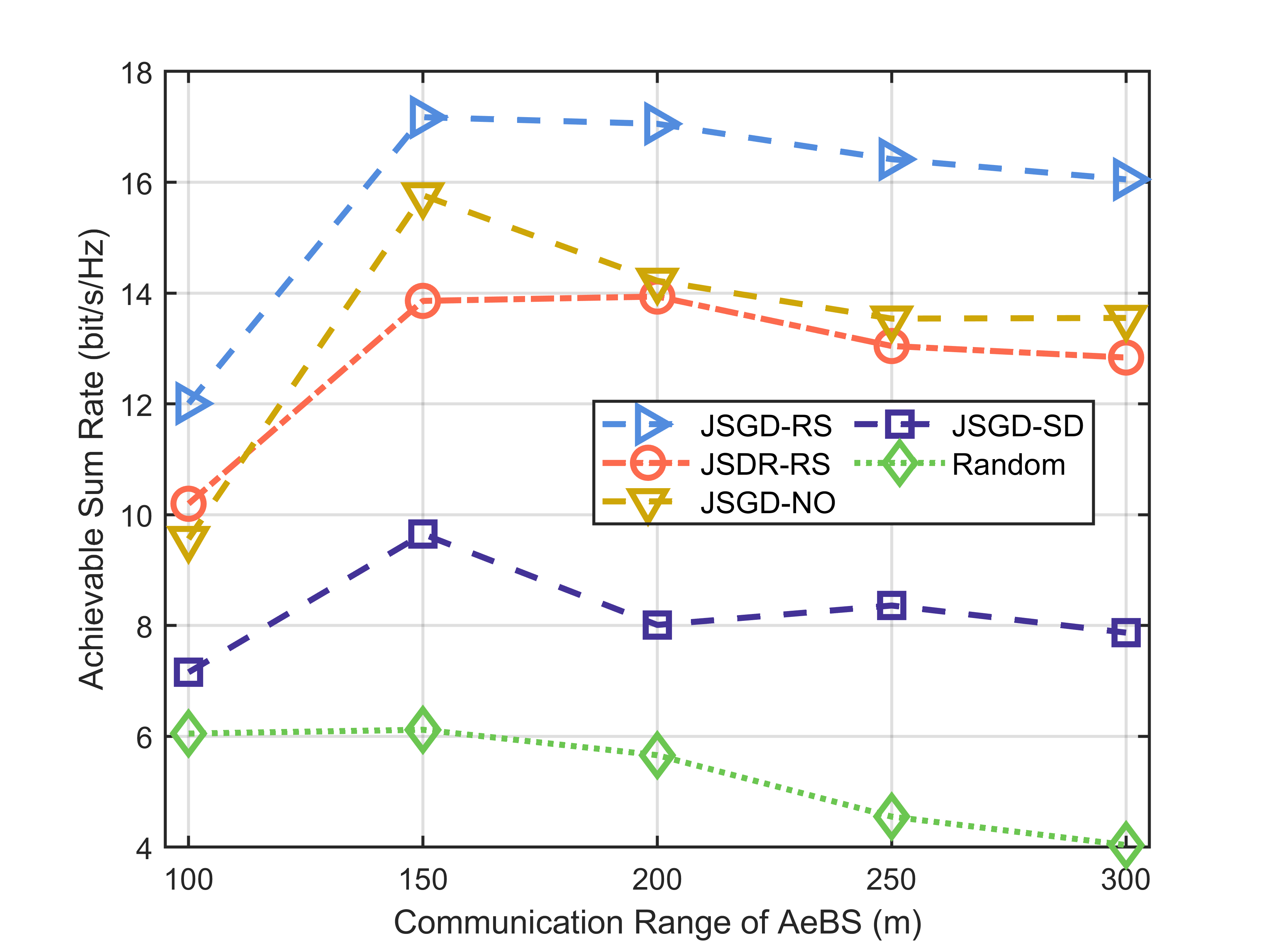}
    \label{sum_rate_vs_communication_range}
}
\subfloat[]{
    \includegraphics[width=0.32\linewidth]{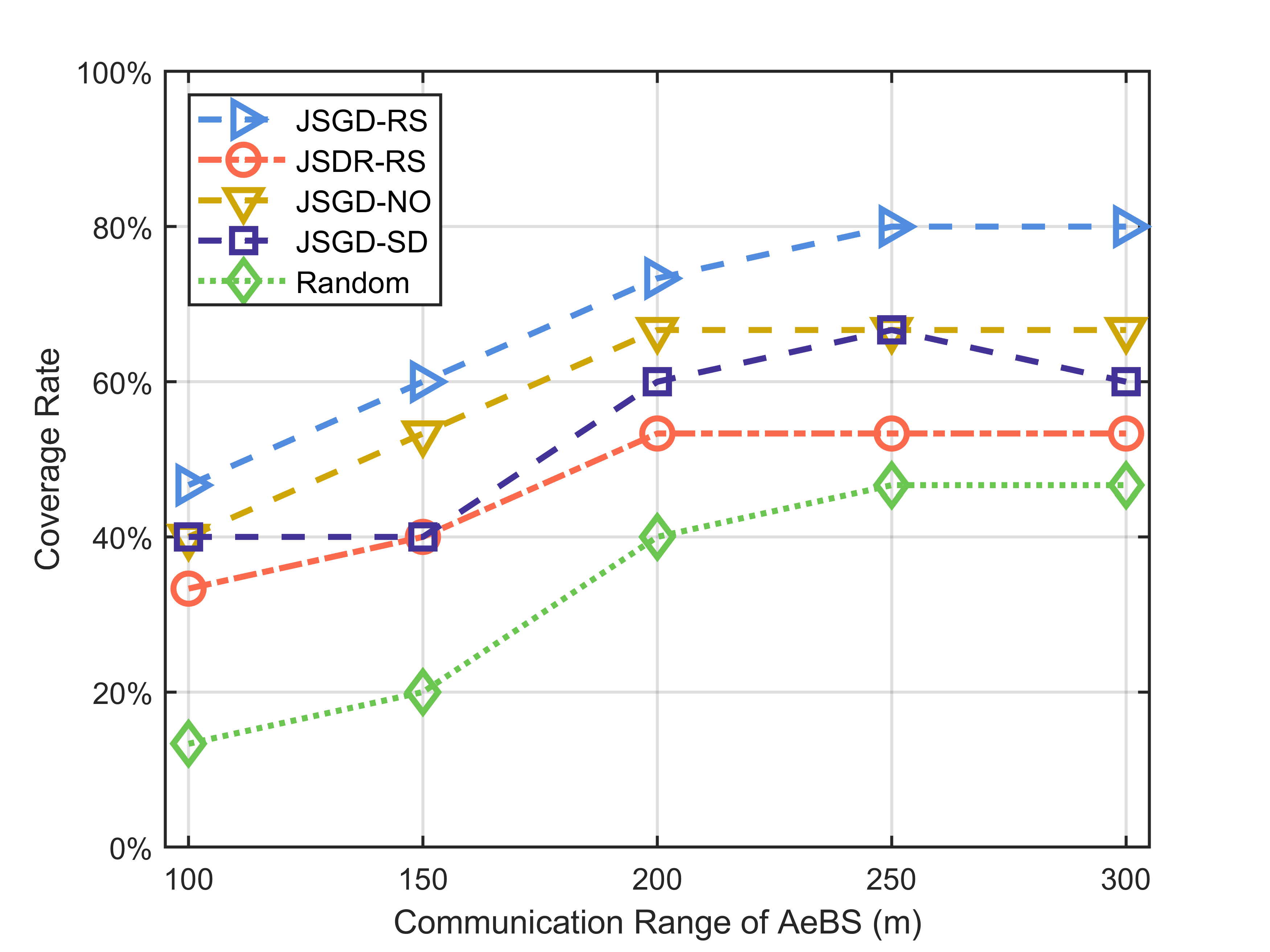}
    \label{coverage_vs_communication_range}
}
\caption{Impact of the communication range of AeBS on: (a) system utility, (b) achievable sum rate, (c) coverage rate.}
\label{impact_of_the_communication_range_of_AeBS}
\end{figure*}

Fig.~\ref{impact_of_number_of_AeBS} shows the impact of the number of AeBSs on system utility, achievable sum rate and coverage rate. In Fig.~\ref{impact_of_number_of_AeBS}(a), as the number of AeBSs increases, all methods exhibit a general upward trend in system utility. This is because additional AeBSs provide enhanced coverage and improved spatial diversity. The proposed JSGD-RS method consistently outperforms all other approaches across different AeBS quantities, demonstrating the effectiveness of combining the RSMA framework with graph diffusion-driven UAV deployment and user association strategies. 
Fig.~\ref{impact_of_number_of_AeBS}(b) presents the relationship between the achievable sum rate and the number of AeBSs for different schemes. First, as the number of AeBSs increases, the achievable sum rate improves for all methods except the Random method. This is because deploying more AeBSs allows more GUs to be served. Second, the proposed JSGD-RS outperforms both JSGD-NO and JSGD-SD. This result indicates that the RSMA framework efficiently manages interference among GUs by leveraging superimposed common data streams and dynamically adjusting rate allocation and beamforming. Moreover, JSGD-RS outperforms the JSDR-RS, confirming the benefits of the generative diffusion-based deployment strategy. Fig.~\ref{impact_of_number_of_AeBS}(c) depicts the coverage rate versus the numbers of AeBSs. Since a greater number of AeBSs offers better coverage capacity, the coverage rate improves as AeBSs increases. However, JSGD-SD achieves decent coverage at lower user counts but its coverage fraction drops more significantly at high user density, reflecting an inability to serve the extra users. Notably, the gap between JSGD-RS and JSDR-RS widens as $K$ increases, indicating that the global graph diffusion approach yields superior coverage, particularly in larger and more complex network environments where effective AeBS positioning and user association become increasingly critical.

Fig.~\ref{impact_of_number_of_GU}(a) shows system utility versus the number of GUs. By jointly leveraging continuous convex approximation for beamforming and rate-splitting, together with graph diffusion-based AeBS deployment and user association, JSGD-RS achieves the highest utility in terms of sum rate and coverage. Fig.~\ref{impact_of_number_of_GU}(b) shows that the achievable sum rate generally increases with more GUs for all schemes except the random baseline. This is because a higher number of GUs introduces more communication links, which can be leveraged effectively by optimized resource allocation and deployment strategies. 
Besides, the proposed JSGD-RS method achieves the highest achievable sum rate, and maintains consistent growth as the number of GUs increases from 10 to 18, indicating its strong scalability and ability to efficiently allocate resources in denser user environments. As the number of GUs increases, the rate performance of JSGD-NO surpasses that of JSDR-RS, indicating that the performance gain brought by the generative diffusion architecture-based deployment strategy becomes dominant in dense user scenarios. Additionally, the rate performance of JSGD-SD shows limited improvement as GUs increases. This limitation arises from the limited number of transmit antennas on each AeBS reduces the spatial separation of beams. Fig.~\ref{impact_of_number_of_GU}(c) shows that JSGD achieves the highest coverage rate, remaining above $70\%$ across all cases. This benefit stems from effective spatial modeling via graph diffusion, enabling optimized AeBS placement. As GUs increases, the overall coverage rate shows a slight downward trend because of the increasing difficulty in satisfying coverage requirements in denser user environments. Notably, the decline in coverage rate for JSGD methods is minimal compared to JSDR, demonstrating the robustness of the proposed method in maintaining efficient coverage in complex network conditions.

Fig.~\ref{impact_of_the_communication_range_of_AeBS}(a) shows that system utility initially increases as the AeBS communication range grows, with JSGD-RS attaining the highest utility across all ranges.  However, the growth rate of system utility gradually decreases with the expansion of the communication range, even the JSDR-RS method shows a decline when the communication range extends from $250~\text{m}$ to $300~\text{m}$. This occurs because, as the communication range reaches medium to large levels, the coverage rate tends to stabilize, while the intensified interference may reduce the achievable sum rate and finally result in slower growth or even a decline. Fig.~\ref{impact_of_the_communication_range_of_AeBS}(b) shows that the achievable sum rate improves for all methods as the communication range increases up to roughly $150~\text{m}$, since a larger communication range allows more GUs to access the LAWNs. However, beyond $150~\text{m}$, the performance of most methods slightly decreases due to increased interference in larger coverage areas. 
Meanwhile, the gap between JSGD-RS and baselines increases, highlighting its superior scalability. Fig.~\ref{impact_of_the_communication_range_of_AeBS}(c) shows that coverage rate increases for both JSGD and JSDR as the range grows from $100~\text{m}$ to $250~\text{m}$, while JSGD consistently outperforms JSDR. This gain is attributed to the graph diffusion model, which effectively captures spatial correlations and environmental structure for improved AeBS placement.

\section{Conclusion}\label{sec_conclusion}

In this paper, we have proposed a novel GenAI-based joint AeBS deployment, user association, and resource allocation strategy in RSMA-enabled LAWNs for supporting URLLC services, thereby maximizing system sum rate and coverage. Specifically, we have developed a graph diffusion model-based algorithm that generated discrete graphs via denoising diffusion to determine the optimal AeBS deployment and user association. Furthermore, we have adopted SCA method to obtain the optimal beamforming and rate allocation strategies for AeBSs. Extensive simulation results have confirmed the effectiveness of our proposed method. Future work will explore LAWNs with practical control stability considerations.



\bibliographystyle{IEEEtran}
\bibliography{references,references_r1}


\end{document}